\documentclass[twocolumn]{aastex631}

\DeclareUnicodeCharacter{2212}{-}

\shorttitle{The PASTA Survey}
\shortauthors{Sun et al.}

\begin{document}
	
	\title{Planets Around Solar Twins/Analogs (PASTA) I.: High precision stellar chemical abundance for 17 planet-hosting stars and the condensation temperature trend\footnote{This paper includes data gathered with the 6.5 meter Magellan Telescopes located at Las Campanas Observatory, Chile.}}
	
	\correspondingauthor{Qinghui Sun, Sharon X. Wang}
	\email{qinghuisun@sjtu.edu.cn, sharonw@tsinghua.edu.cn}
	
	\author[0000-0003-3281-6461]{Qinghui Sun}
	\affiliation{Tsung-Dao Lee Institute, Shanghai Jiao Tong University, Shanghai, 200240, China}
	\affiliation{Department of Astronomy, Tsinghua University, Beijing, 100084, China}
	
	\author[0000-0002-6937-9034]{Sharon Xuesong Wang}
	\affiliation{Department of Astronomy, Tsinghua University, Beijing, 100084, China}
	
	\author[0000-0002-4503-9705]{Tianjun Gan}
	\affiliation{Department of Astronomy, Tsinghua University, Beijing, 100084, China}
	
	\author[0009-0005-3557-5183]{Chenyang Ji}
	\affiliation{Department of Astronomy, Tsinghua University, Beijing, 100084, China}
	
	\author{Zitao Lin}
	\affiliation{Department of Astronomy, Tsinghua University, Beijing, 100084, China}
	
	\author[0000-0001-5082-9536]{Yuan-Sen Ting}
	\affiliation{Department of Astronomy, The Ohio State University, 1251 Wescoe Hall Dr., Columbus, Ohio, 43210, USA}
	\affiliation{Center for Cosmology and AstroParticle Physics, The Ohio State University, 191 West Woodruff Avenue, Columbus, Ohio, 43210, USA}
	\affiliation{Max Planck Institute for Astronomy, Königstuhl 17, D-69117 Heidelberg, Germany}
	
	\author[0009-0008-2801-5040]{Johanna Teske}
	\affiliation{Earth and Planets Laboratory, Carnegie Institution for Science, 5241 Broad Branch Road, NW, Washington, DC 20015, USA}
	
	\author{Haining Li}
	\affiliation{Key Lab of Optical Astronomy, National Astronomical Observatories, Chinese Academy of Sciences (CAS) A20 Datun Road, Chaoyang, Beijing 100101, People's Republic of China}
	
	\author{Fan Liu}
	\affiliation{School of Physics and Astronomy, Monash University, Melbourne, VIC, 3800, Australia}
	\affiliation{ARC Centre of Excellence for Astrophysics in Three Dimensions (ASTRO-3D), Canberra, ACT, 2611, Australia}
	
	\author[0000-0001-7916-4371]{Xinyan Hua}
	\affiliation{Department of Astronomy, Tsinghua University, Beijing, 100084, China}
	
	\author{Jiaxin Tang}
	\affiliation{Department of Astronomy, Tsinghua University, Beijing, 100084, China}
	
	\author[0000-0002-0007-6211]{Jie Yu}
	\affiliation{School of Computing, Australian National University, Acton, ACT 2601, Australia}
	\affiliation{Research School of Astronomy \& Astrophysics, Australian National University, Cotter Rd., Weston, ACT 2611, Australia}
	
	\author{Jiayue Zhang}
	\affiliation{Department of Astronomy, Tsinghua University, Beijing, 100084, China}
	
	\author[0000-0003-4903-567X]{Mariona Badenas-Agusti}
	\affiliation{Department of Earth, Atmospheric and Planetary Sciences, Massachusetts Institute of Technology, Cambridge, MA 02139, USA}
	\affiliation{Department of Physics and Kavli Institute for Astrophysics and Space Research, Massachusetts Institute of Technology, Cambridge, MA 02139, USA}
	\affiliation{MIT William Asbjornsen Albert Memorial Fellow}
	
	\author[0000-0001-7246-5438]{Andrew Vanderburg}
	\affiliation{Department of Physics and Kavli Institute for Astrophysics and Space Research, Massachusetts Institute of Technology, Cambridge, MA 02139, USA}
	
	\author[0000-0003-2058-6662]{George~R.~Ricker}
	\affil{Department of Physics and Kavli Institute for Astrophysics and Space Research, Massachusetts Institute of Technology, Cambridge, MA 02139, USA}
	
	\author[0000-0001-6763-6562]{Roland~Vanderspek}
	\affil{Department of Physics and Kavli Institute for Astrophysics and Space Research, Massachusetts Institute of Technology, Cambridge, MA 02139, USA}
	
	\author[0000-0001-9911-7388]{David~W.~Latham}
	\affil{Center for Astrophysics ${\rm \mid}$ Harvard {\rm \&} Smithsonian, 60 Garden Street, Cambridge, MA 02138, USA}
	
	\author[0000-0002-6892-6948]{Sara~Seager}
	\affil{Department of Physics and Kavli Institute for Astrophysics and Space Research, Massachusetts Institute of Technology, Cambridge, MA 02139, USA}
	\affil{Department of Earth, Atmospheric and Planetary Science, Massachusetts Institute of Technology, 77 Massachusetts Avenue, Cambridge, MA 02139, USA}
	\affil{Department of Aeronautics and Astronautics, MIT, 77 Massachusetts Avenue, Cambridge, MA 02139, USA}
	
	\author[0000-0002-4715-9460]{Jon~M.~Jenkins}
	\affil{NASA Ames Research Center, Moffett Field, CA 94035, USA}
	
	\author[0000-0001-8227-1020]{Richard P. Schwarz}
	\affil{Center for Astrophysics ${\rm \mid}$ Harvard {\rm \&} Smithsonian, 60 Garden Street, Cambridge, MA 02138, USA}
	
	\author[0000-0002-7188-8428]{Tristan Guillot}
	\affil{Observatoire de la C\^ote d'Azur, UniCA, Laboratoire Lagrange, CNRS UMR 7293, CS 34229, 06304 Nice cedex 4, France}
	
	\author[0000-0001-5603-6895]{Thiam-Guan Tan}
	\affiliation{Perth Exoplanet Survey Telescope, Perth, Western Australia}
	
	\author[0000-0003-2239-0567]{Dennis M.\ Conti}
	\affiliation{American Association of Variable Star Observers, 185 Alewife Brook Parkway, Suite 410, Cambridge, MA 02138, USA}
	
	\author[0000-0003-2781-3207]{Kevin I.\ Collins}
	\affiliation{George Mason University, 4400 University Drive, Fairfax, VA, 22030 USA}
	
	\author{Gregor Srdoc}
	\affil{Kotizarovci Observatory, Sarsoni 90, 51216 Viskovo, Croatia}
	
	\author[0000-0003-2163-1437]{Chris Stockdale}
	\affiliation{Hazelwood Observatory, Australia}
	
	\author{Olga Suarez}
	\affiliation{Universit\'e C\^ote d'Azur, Observatoire de la C\^ote d'Azur, CNRS, Laboratoire Lagrange, Bd de l'Observatoire, CS 34229, 06304 Nice cedex 4, France}
	
	\author{Roberto Zambelli} 
	\affiliation{Società Astronomica Lunae, Castelnuovo Magra, Italy}
	
	\author[0000-0002-3940-2360]{Don Radford} 
	\affiliation{Brierfield Observatory, Bowral, NSW Australia}
	
	\author[0000-0003-1464-9276]{Khalid Barkaoui}
	\affiliation{Astrobiology Research Unit, Universit\'e de Li\`ege, 19C All\'ee du 6 Ao\^ut, 4000 Li\`ege, Belgium}
	\affiliation{Department of Earth, Atmospheric and Planetary Science, Massachusetts Institute of Technology, 77 Massachusetts Avenue, Cambridge, MA 02139, USA}
	\affiliation{Instituto de Astrof\'isica de Canarias (IAC), Calle V\'ia L\'actea s/n, 38200, La Laguna, Tenerife, Spain}
	
	\author[0000-0002-5674-2404]{Phil Evans}
	\affiliation{El Sauce Observatory, Coquimbo Province, Chile}
	
	\author[0000-0001-6637-5401]{Allyson Bieryla}
	\affil{Center for Astrophysics ${\rm \mid}$ Harvard {\rm \&} Smithsonian, 60 Garden Street, Cambridge, MA 02138, USA}
	
	\begin{abstract}
		
		The Sun is depleted in refractory elements compared to nearby solar twins, which may be linked to the formation of giant or terrestrial planets. Here we present high-resolution, high signal-to-noise spectroscopic data for 17 solar-like stars hosting planets, obtained with Magellan II/MIKE, to investigate whether this depletion is related to planet formation. We derive stellar parameters, including stellar atmosphere, age, radius, mass, and chemical abundances for 22 elements from carbon to europium through line-by-line differential analysis. Our uncertainties range from 0.01 dex for Fe and Si to 0.08 dex for Sr, Y, and Eu. By comparing the solar abundances to those of the 17 stars, we investigate the differential abundance ([X/Fe]$_{\rm solar}$ - [X/Fe]$_{\rm star}$) versus condensation temperature ($T_c$) trend. In particular, we apply Galactic chemical evolution corrections to five solar twins within the full sample. Our results conform to previous studies that the Sun is relatively depleted in refractory compared to volatile elements. For both five solar twins and the rest of solar-like stars, we find that all stars hosting known gas giant planets exhibit negative $T_c$ trend slopes, suggesting that the Sun is relatively depleted in refractory elements compared to similar giant-planet-host stars. Additionally, we find no correlation between $T_c$ trend slopes and the total mass of detected terrestrial planets in each system, suggesting that terrestrial planet formation may not be the cause of refractory element depletion in the Sun.
		
	\end{abstract}
	
	\section{Introduction} \label{sec:intro}
	
	The Sun serves as a fundamental benchmark for studying star formation and calibrating stellar evolution models. Previous research has extensively explored various aspects of solar compositions (e.g., \citealt{2005ARA&A..43..481A, 2005ApJ...621L..85B, 2009ARA&A..47..481A, 2021SSRv..217...44L, 2024arXiv240902103T}), despite ongoing debates about its chemical makeup. Of particular interest are the compositions of volatile versus refractory elements, which are categorized based on their condensation temperatures ($T_c$) in the protoplanetary disk. Refractory elements, such as iron and magnesium, condense at higher temperatures, while volatile elements like sodium and potassium remain gaseous under cooler conditions. These differences are supposed to play a significant role in the chemical composition of planets that form from the surrounding material of their host stars.
	
	Solar-like stars are broadly similar to the Sun, with solar analogs showing similar mass, temperature, and metallicity, while solar twins have nearly identical parameters. \citet{Bedell2018} proposed a definition for solar twins, specifying that their effective temperature differs from the Sun by less than 100 K, their surface gravity also differs by less than 100 K, and their metallicity differs by less than 0.1 dex. Solar twins are valuable for determining the uniqueness of the Sun and our planetary system.
	
	\citet{2009ApJ...704L..66M} first noted that our Sun shows a depletion of refractory elements relative to volatile elements when compared to solar twins, based on the differential elemental abundance versus condensation temperature ($T_c$) trend. This observation has been supported by more recent studies (e.g., \citealt{2021ApJ...907..116N, 2024arXiv240216954R}). It has been proposed that this may be explained by the retention of refractory elements into small planets. The mass needed was estimated by \citet{2010ApJ...724...92C} to amount to approximately four Earth-like planets, assuming a 50/50 mix of Earth-like and Cometary-like chondrite material. However, this value did not account for stellar evolution and the fact that the solar convective zone was at least one order of magnitude more massive when the solar system formed. A self-consistent calculation accounting for mixing in the solar interior showed that explaining the lack of refractory elements requires Jupiter, Saturn, Uranus and Neptune to also have accreted more rocky materials than icy materials \citet{2018A&A...618A.132K}.
	
	The formation of planets must, to some extent, affect stellar compositions. \citet{2020MNRAS.493.5079B} proposed that when a forming giant planet forms inside the snow line, it opens a gap in the gas disk and creates a pressure bump that can trap over 100 Earth masses of dust outside its orbit. This prevents this dust from accreting onto the star and may result in a refractory elements depletion of $\sim$ 5-15\% in the host star. Moreover, recent work by \citet{2021A&A...655A..51K} suggests that this process may impact the composition of the Sun’s core, potentially offering a more accurate explanation of the observed solar neutrino flux \citep{2022A&A...667L...2K}. Differences in volatile-to-refractory ratios have also been linked to the origins of close-in giant planets, with higher refractory-to-volatile ratios predicted under the planet migration scenario compared to the in-situ formation scenario (\citealt{2022A&A...665L...5K}). However, no observational evidence has yet supported this, partly due to the limited availability of high-quality spectra for solar twins hosting giant planets.
	
	Alternatively, a $T_c$ trend like that observed in \citet{2014ApJ...791...14M} may instead indicate that the early solar nebula experienced significant material loss due to processes such as gas dispersion and substantial heating during accretion events (\citealt{2014A&A...562A..27T}). Alternatively, this trend might suggest late-stage accretion of planetary material (e.g., \citealt{2011ApJ...740...76R, 2015A&A...582L...6S, 2018ApJ...854..138O}), implying that some solar analogs may have undergone planet engulfment events in the past. Furthermore, the trapping of refractory material by gas giant planets could also contribute to a positive $T_c$ trend slope (\citealt{2020MNRAS.493.5079B}).
	
	The $T_c$ trends are also influenced by galactic chemical evolution (GCE), where stars formed at different locations from the galactic center with varying ages and metallicities can naturally impact the $T_c$ trend (e.g. \citealt{Bedell2018}), irrespective of planet formation. Therefore, we need precise stellar parameter measurements such as age to perform the GCE correction. Also given that the $T_c$ trend analysis requires high precision elemental abundance across a range of condensation temperatures, high-resolution, high signal-to-noise (S/N) ratio, and broad wavelength coverage spectra are needed.
	
	Though various studies have been conducted and different explanations proposed, it remains unclear whether the Sun is normal in its volatile versus refractory element compositions. The existing sample of solar twins and analogs with well-measured abundances is primarily composed of non-planet-hosting stars, while the sample of planet-hosting solar analogs/twins remains small. For example, \citet{2010ApJ...720.1592G} used high-resolution, high S/N spectra (R $\geq$ 85,000, S/N $\sim$ 800) to derive high-precision (0.02 –- 0.03 dex) elemental abundances for seven solar twins and 95 solar analogs, including 24 planet-hosting stars and 71 stars without detected planets. Additionally, \citet{Bedell2018} studied 79 solar-like stars (spectra R = 83,000 –- 65,000, S/N $\sim$ 400), 68 of which are classified as solar twins, four of which host planets. \citet{2009ApJ...704L..66M} focused on 11 solar twins and 10 solar analogs (spectra R = 65,000, S/N $\sim$ 450, $\sigma$[X/Fe] $\sim$ 0.01 dex), with four solar analogs hosting planets. Furthermore, \citet{2015A&A...579A..52N, 2016A&A...593A..65N} performed high-precision abundance analyses (R $\sim$ 115,000, S/N $>$ 600, $\sigma$[X/Fe] $\sim$ 0.01 dex) of 21 solar twins, three of which are known to host planets. Despite these efforts, the number of solar twins hosting planets remains quite small (fewer than ten), and the number of solar analogs hosting planets is similarly limited at around 40. Moreover, the variety of abundance analysis methods and instruments used across studies often leads to systematic shifts, complicating direct comparisons between them. Key questions persist: Does the $T_c$ slope of the Sun differ from those solar twins? Are the $T_c$ slopes of solar twins consistent among themselves? Do the $T_c$ slopes of planet-hosting solar twins differ from those of non-planet-hosting ones? Do the $T_c$ trend slopes of stars hosting giant planets differ from those hosting small planets? Do the $T_c$ trend slopes depend on planet mass? The $T_c$ trend is crucial for exploring potential planetary formation scenarios, and for investigating the complex relationship between stellar chemical abundances and various stellar and planetary parameters.
	
	Inspired by the need to understand solar composition and planetary system formation, we initiate the PASTA: Planets Around Solar Twins/Analogs Survey. The primary objective is to acquire high-resolution, high signal-to-noise (S/N) spectra and determine their stellar parameters, including the chemical abundances of elements across a broad range of condensation temperatures. Several efforts have been made to field solar twins/analogs without known planets, which we can use for comparison. This is the first paper of the PASTA survey, which covers 17 solar-like stars hosting confirmed planets or planet candidates. The goal of the full survey is to observe $\sim$ 100 solar twins/analogs hosting planets using high-resolution spectroscopy, allowing for a strict comparison with the Sun to determine if its composition is statistically distinguishable from that of the twins. We aim to address whether the Sun is normal concerning its volatile versus refractory composition. Additionally, the survey aims to investigate how the presence of giant/small planets may influence the $T_c$ slope and the volatile and refractory composition. This survey will expand the existing sample of planet-hosting solar twins/analogs with well-measured high-precision stellar parameters. In addition, even within planet-hosting solar twin studies, there is either no clear distinction between rocky planets and gas giants (e.g., \citealt{2010ApJ...720.1592G}), or the sample lacks one type of planet for comparison (e.g., \citealt{Bedell2018}). While TESS gives us this unique opportunity. In this paper, we discuss target selection and observations in Section \ref{sec:selection}, report high-precision stellar atmospheric parameters and elemental abundances for 22 elements ranging from C to Eu (with an average precision of approximately 0.04 dex) in Section \ref{sec:abundance}, derive stellar isochrone age, mass, and radius in Section \ref{sec:age}, analyze the $T_c$ trend and compare solar twins to the Sun in Sections \ref{sec:GCE} and \ref{sec:Tc}, discuss the potential effect of planet formation on the depletion of refractory elements in Section \ref{sec:discussion}, and provide a summary of the paper in Section \ref{sec:summary}.
	
	\section{Target Selection and Observations}  \label{sec:selection}
	
	We constructed a sample of stars that host known/confirmed planets, or planet candidates by selecting solar-like stars from the TESS Objects of Interest (TOIs) Catalog. According to \cite{Bedell2018}, solar twins are stars with an effective temperature ($T_{\rm eff}$) within 100 K of the solar value, a surface gravity (log {\it g}) within 0.1 dex of solar, and a metallicity ([Fe/H]) within 0.1 dex of solar. For solar analogs, the criteria are slightly broadened to $|\Delta T_{\rm eff}| < 200$ K, $|\Delta$ log {\it g}$|$ $<$ 0.2, and $|\Delta$[Fe/H]$|$ $<$ 0.2 dex compared to the Sun. For initial target selection, we adopt $|\Delta T_{\rm eff}| < 200$ K, $|\Delta$ log {\it g}$|$ $<$ 0.2, and $|\Delta$[Fe/H]$|$ $<$ 0.4 dex to select targets from the TOI catalog. This approach allows us to include a wider range of solar-like stars with varying stellar parameters around the Sun, thereby increasing the chances of identifying solar twins. We select targets covering a wider range of metallicities than solar analogs, also partly to enable comparisons with metal-poor and metal-rich stars of different ages, which are relevant to studies of Galactic Chemical Evolution. Since stellar parameters from the TOI catalog may be imprecise, high-resolution and high S/N spectra are necessary to confidently confirm solar twins, which is one of the goal of this survey. We first identify 188 solar-like stars with known/confirmed planets or planet candidates by cross-matching the TESS Input Catalog (TIC; \citealt{Stassun2018_TIC1, Stassun2019_TIC2, https://doi.org/10.17909/fwdt-2x66}) with the Gaia solar analog catalog (\citealt{GaiaSolar2023}) based on Gaia DR3 ID or coordinates, thereby establishing preliminary constraints based on Gaia measurements of $T_{\rm eff}$, log g, and [Fe/H]. We use the solar analog catalog from Table 7 of \citet{GaiaSolar2023}, which includes 5,863 targets. Within the sample, the stellar parameters from spectroscopy shows an average deviation from solar values of 14.4 K in $T_{\rm eff}$, −0.071 in log g, and −0.05 dex in metallicity ([M/H]). The masses of the stars range from 0.95 M$_{\odot}$ to 1.05 M$_{\odot}$, while their radii vary between 0.8 R$_{\odot}$ and 1.2 R$_{\odot}$. The full list of 188 solar-like stars selected for the PASTA survey is provided in Appendix Table A, with the full table available online in machine-readable format.
	
	We then prioritize our target list based on: the likelihood of being solar twins and the likelihood of hosting planets. For the first factor, we examine the stellar parameters derived from the high-resolution (R $\sim$ 80,000) but low S/N ($\sim$ 10 -- 50) spectra obtained by TESS Follow-up Observing Program Working Group (TFOP-WG) Sub Group 2 (SG2, reconnaissance spectroscopy) available on the Exoplanet Follow-up Observing Program (ExoFOP) website\footnote{\url{https://exofop.ipac.caltech.edu/tess/}} (\citealt{https://doi.org/10.26134/exofop1}). If any parameter exceeds the range specified by \cite{Bedell2018} over 2-$\sigma$ error, we downgrade the priority of the target. For the second factor, we check on the TESS light curves and the disposition from TFOP-WG Sub Group 1 (SG1), which relies primarily on ground-based follow-up observations. We then assess the likelihood that the planet candidates are genuine, exclude potential eclipsing binaries and those with weak transit signals, and prioritize the target list accordingly. For the first paper of PASTA, we observe 17 solar-like planet-host stars by using the above selection process. The 17 targets are prioritized on our list and can be observed using Magellan II/MIKE during the observing period. The three targets observed in 2022 were a preliminary assessment of the resolution and S/N necessary for this project. For the 2024A observing run, we made slight improvements to both the resolution and S/N, enabling us to observe 14 targets throughout the entire night. In the upcoming December 2024B Magellan II/MIKE observing run (one night), we plan to cover an additional $\sim$ 15 Southern targets from our list that are observable with Magellan II/MIKE, completing our observations of Southern targets. For the Northern targets, we currently have 20 hours allocated on LBT/PEPSI for the 2025A observing run.
	
\begin{splitdeluxetable*}{cccp{1.0cm}p{1.0cm}p{0.9cm}ccccBccccccccccc}
	\label{tab:parameters}
	\tablecaption{The Observing Log and Target Parameters}
	\tabletypesize{\scriptsize}
	\tablehead{
		\colhead{TOI ID$^1$} & \colhead{UT Date$^2$} & \colhead{$T$mag$^1$} & \colhead{$P_{orb}^1$}	& \colhead{$R_p^1$}	& \colhead{$T_{eq}^1$} & \colhead{WG$^1$} & \colhead{FPP$^1$} & \colhead{NFPP$^1$} &
		\colhead{$T_{\rm eff}^3$} & \colhead{$\sigma_{Teff}^3$} & \colhead{log $g^3$} & \colhead{$\sigma_{logg}^3$} & \colhead{$V_t^3$} & \colhead{$\sigma_{V_t}^3$} & \colhead{age$^4$} & \colhead{$\sigma_{age}^4$} & \colhead{$R_{\ast}^4$} & \colhead{$\sigma_{R_{\ast}}^4$} & \colhead{$M_{\ast}^4$} & \colhead{$\sigma_{M_{\ast}}^4$} \\
		\colhead{} & \colhead{} & \colhead{} & \colhead{day}	& \colhead{$R_{\bigoplus}$}	& \colhead{K} & \colhead{} & \colhead{} & \colhead{} &
		\colhead{K} & \colhead{K} & \colhead{} & \colhead{} & \colhead{km s$^{-1}$} & \colhead{km s$^{-1}$} & \colhead{Gyr} & \colhead{Gyr} & \colhead{$R_{\odot}$} & \colhead{$R_{\odot}$} & \colhead{$M_{\odot}$} & \colhead{$M_{\odot}$}
	} 
	\startdata
	\hline
	TOI-1036 &	2024-05-04 &	10.286 &	              3.780	&               5.984 &	          1294 & VPC	 & 0.061 & 0.049   & 5761	& 20 &	4.37 & 0.05	& 0.77 & 0.06 &	 4.444 & 1.132 & 1.114	& 0.075	& 1.072	& 0.016 \\
	TOI-1055 &	2024-05-04 &	 8.089 &	             17.471	&               3.165 &	           781 & CP	& -- & --    & 5769	& 18 &	4.35 & 0.05	& 0.85 & 0.05 &	 6.959 & 1.235 & 1.095	& 0.066	& 0.994	& 0.010 \\
	TOI-1076 &	2024-05-04 &	11.247 &	              2.550	&              12.888 &	          1338 & KP$^a$	& -- & --    & 5719	& 32 &	4.53 & 0.05	& 0.64 & 0.11 &	 1.577 & 0.966 & 1.020	& 0.028	& 1.093	& 0.014 \\
	TOI-1097 &	2024-05-04 &	 8.722 &	      9.189, 13.903	&        2.174, 2.614 &	      811, 757 & VP	&  NaN  &  NaN & 5935	& 37 &	4.59 & 0.07	& 1.23 & 0.08 &	 1.737 & 1.135 & 1.003	& 0.034	& 1.063	& 0.018 \\
	TOI-1117 &	2022-08-29 &	10.406 &	              2.228	&               2.568 &	          1434 & CP$^g$	&  --  &  --  & 5630	& 19 &	4.40 & 0.05	& 0.62 & 0.08 &	 5.876 & 1.641 & 1.055	& 0.059	& 1.012	& 0.012 \\
	TOI-1203 &	2024-05-04 &	 8.000 &	              2.520	&               2.964 &	           663 & VP	&  NaN  &  NaN      & 5724	& 29 &	4.17 & 0.07	& 0.88 & 0.06 &	12.482 & 1.028 & 1.286	& 0.126	& 0.896	& 0.029 \\
	TOI-215 &	2022-08-31 &	10.501 &	             26.285	&               3.083 &	           592 & VPC	&  NaN  &   NaN     & 5757	& 39 &	4.56 & 0.06	& 0.76 & 0.10 &	 1.881 & 1.203 & 0.988	& 0.031	& 1.058	& 0.018 \\
	TOI-2011 &	2024-05-04 &	 5.049 &	11.580, 27.590, NaN	& 1.542, 2.705, 1.621 &	806, 604, 1092 & KP$^b$	& --  & --       & 5751	& 19 &	4.49 & 0.05	& 0.85 & 0.05 &	 5.489 & 2.177 & 0.906	& 0.039	& 0.913	& 0.015 \\
	TOI-2426 &	2022-08-30 &	 9.970 &	              4.082	&               2.382 &	          1226 & CPC	& NaN &  NaN      & 5705	& 12 &	4.37 & 0.03	& 0.84 & 0.03 &	 7.881 & 0.852 & 1.062	& 0.037	& 0.966	& 0.006 \\
	TOI-3342 &	2024-05-04 &	11.848 &	              5.971	&              12.337 &	          1021 & CP$^g$ &  --  & --      & 6029	& 30 &	4.57 & 0.05	& 1.25 & 0.06 &	 0.865 & 0.578 & 1.065	& 0.020	& 1.141	& 0.014 \\
	TOI-4628 &	2024-05-04 &	 9.666 &	              6.327	&               2.290 &	           954 & KP$^c$	&  --  &  --      & 5810	& 46 &	4.53 & 0.07	& 0.81 & 0.10 &	 2.466 & 1.496 & 1.004	& 0.044	& 1.051	& 0.022 \\
	TOI-4914 &	2024-05-04 &	11.663 &	             10.599	&              12.605 &	           574 & CP$^g$	& --   &  --      & 5943	& 43 &	4.55 & 0.08	& 0.83 & 0.10 &	 2.262 & 1.384 & 1.020	& 0.047	& 1.061	& 0.019 \\
	TOI-5005 &	2024-05-04 &	11.130 &	              6.308	&               6.004 &	          1012 & CP$^g$	&  --  &  --      & 5704	& 17 &	4.48 & 0.05	& 0.59 & 0.09 &	 2.804 & 1.286 & 1.007	& 0.037	& 1.051	& 0.011 \\
	TOI-5795 &	2024-05-04 &	10.187 &	              6.141	&               5.427 &	          1005 & VPC	& 0.0185   &  0.000      & 5697	& 26 &	4.27 & 0.07	& 0.60 & 0.07 &	11.081 & 1.171 & 1.148	& 0.101	& 0.919	& 0.016 \\
	TOI-744 &	2024-05-04 &	 9.757 &	              4.318	&              13.393 &	          1084 & KP$^d$	&  --  &  --      & 5721	& 17 &	4.42 & 0.05	& 0.87 & 0.05 &	 5.924 & 1.801 & 1.002	& 0.050	& 0.973	& 0.012 \\
	TOI-755 &	2024-05-04 &	 9.463 &	       2.540, 6.742	&        2.153, 2.114 &	     1391, 952 & KP$^e$ &  --  &  --  	& 5730	& 19 &	4.47 & 0.06	& 0.66 & 0.08 &	 3.947 & 1.773 & 0.996	& 0.048	& 1.012	& 0.013 \\
	TOI-818 &	2024-05-04 &	10.648 &	              3.119	&              11.857 &	          1195 & KP$^f$  &  --  &  --	& 5737	& 17 &	4.44 & 0.06	& 0.71 & 0.07 &	 4.623 & 1.789 & 1.014	& 0.053	& 1.008	& 0.013 \\
	\hline
	\enddata
	\tablecomments{1}{Tess Object of Interest (TOI) Catalog (\citealt{2021ApJS..254...39G}) IDs, Tess magnitude ($T$mag), orbital period ($P_{orb}$), planet radius ($R_p$), equilibrium temperature ($T_{eq}$), and TESS Follow-up Observing Program (TFOP) Working Group (WG) designation \footnote{\url{https://tess.mit.edu/followup/}}. All these information are directly adopted from the TOI catalog. The WG designation ``CP" means confirmed planet, ``KP" means known planet, ``CPC" means disposition cleared planetary candidate (CPC), ``VPC" means verified planet candidate, VPC+ means verified achromatic planet candidate, ``VP" means validated planet. If a TOI star hosts multiple planets, the parameters for each planet are separated by ``," or ``/". FPP and NFPP are the false positive probability and nearby false positive probability, computed by using the TRICERATOPS code (\citealt{2020ascl.soft02004G}).}
	\tablecomments{2}{UT date when observations began.} 
	\tablecomments{3}{Effective temperature ($T_{\rm eff}$), surface gravity (log {\it g}), microturbulence ($V_t$) and their associated errors derived from this work.}
	\tablecomments{4}{We derive the age, stellar radius, and stellar mass by using the q$^2$ package (\citealt{Ramirez2014}) and the Yonsei-Yale (Y$^2$) isochrones (\citealt{2001ApJS..136..417Y}).}
	\tablecomments{5}{a. TOI-1076 (HATS-33, \citealt{2016AJ....152..161D}); b. TOI-2011 (HD 136352, \citealt{2019AA...622A..37U}); c. TOI-4628 (K2-277, \citealt{2018AJ....156..277L}); d. TOI-744 (WASP-34, \citealt{2011AA...526A.130S}); e. TOI-755 (HD 110113, \citealt{2021MNRAS.502.4842O}; f. TOI-818 (WASP-16, \citealt{2009ApJ...703..752L}); g. These planets have been recently confirmed by the TFOP SG1 group but the discovery paper have not been published yet.}
	\vspace{-10mm}
\end{splitdeluxetable*}
	
	In Table \ref{tab:parameters}, we present the TESS Object of Interest (TOI, \citealt{2021ApJS..254...39G}) Catalog ID, the planet's orbital period, radius, equilibrium temperature, and the TESS Follow-up Observing Program (TFOP) Working Group (WG) designation for each planet, all retrieved directly from the NASA Exoplanet Archive (\citealt{https://doi.org/10.26133/nea1}). For planet candidates with the Science Processing Operations Center (SPOC; \citealt{jenkinsSPOC2016}; located at the NASA Ames Research Center) or TESS-SPOC (\citealt{TESSSPOC2020}) SAP (\citealt{twicken:PA2010SPIE, morris:PA2020KDPH}) light curves available, we calculate the false positive probability (FPP) and nearby false positive probability (NFPP) using the TRICERATOPS code (\citealt{2021AJ....161...24G, 2020ascl.soft02004G}). The signal-to-noise ratio (S/N) of the light curves for TOI-1097, TOI-1203, TOI-215, and TOI-2426 is too low to compute the FPP and NFPP. Based on \citet{2021AJ....161...24G}, the criteria for classifying TOIs as validated planets need to satisfy FPP $<$ 0.015 and NFPP $<$ 10$^{-3}$ , while likely planets satisfy FPP $<$ 0.5 and NFPP $<$ 10$^{-3}$. TOI-1036 and TOI-5795 are likely planets. All other TOIs in our sample are known (published) or confirmed (by TESS but not yet published) planets.
	
	We obtained high S/N ratio, high resolution ($R$) spectroscopy data for the 17 targets by using the Double Echelle Spectrograph for the Magellan Telescopes (MIKE, \citealt{2003SPIE.4841.1694B}). The quartz flats, milky flats and fringe flats were taken at the beginning of the night. Multiple ThAr lamp spectra were taken regularly before and after the object spectra exposures. For the three stars observed in 2022, spectra were obtained with a 0.7" slit, resulting in a resolution of $\sim$ 32,000 on the red side and $\sim$ 41,000 on the blue side, with S/N values above 100 near 7000 \AA . In the 2024 observing run, all spectra were taken using the 0.5" slit to achieve higher resolution than the 2022 run, achieving a resolution of $\sim$ 44,000 on the red side and $\sim$ 57,000 on the blue side, with a signal-to-noise ratio (S/N) exceeding 200 near 7000 \AA\ for all observed stars. All spectra cover a wavelength range of 3350-5000 \AA\ in blue and 4900-9500\AA\ in red. 
	
	The data were processed by using the Carpy standard pipeline\footnote{https://github.com/alexji/mikerun} (\citealt{2000ApJ...531..159K, 2003PASP..115..688K}). This pipeline automatically handles tasks including dealing with the overscan region, flats, wavelength calibration, and combining the spectra with the same object name into red- and blue-side spectra. We then normalize the continuum of the combined spectra to uniform for each star.
	
	\section{Stellar atmosphere and abundance}  \label{sec:abundance}
	
	\subsection{Equivalent Width and Stellar Atmosphere}
	
	The linelist used in this study was adopted from Table 1 of \citet{2014ApJ...791...14M}. For elements with atomic numbers $\le$ 30, the equivalent widths (EWs) were measured by using the \textit{splot} task in IRAF \footnote[5]{IRAF is distributed by the National Optical Astronomy Observatories, operated by the Association of Universities for Research in Astronomy Inc., under a cooperative agreement with the National Science Foundation.}. We performed a line-by-line differential analysis by comparing the object spectra with the solar spectrum (asteroid Vesta) obtained also by MIKE/Magellan. We present the wavelength, element number, excitation status, excitation potential, log(gf) value, and damping C for each line, along with the equivalent widths (EWs) for each line for each star, including the Sun, in Table \ref{tab:EW}. We employed the q$^2$ Python package (\citealt{Ramirez2014}) to derive stellar atmosphere parameters, which include the effective temperature ($T_{\rm eff}$), surface gravity (log $g$), microturbulence ($V_t$), and metallicity ([Fe/H]\footnote{[X/H] = A(X)$_{star}$ - A(X)$_{Sun}$, where X is the element species. A(X) = 12 + log(N$_X$/N$_H$), where N$_X$ is number of atoms of species X.}). In this paper we adopt $T_{\rm eff}$ = 5777 K, log {\it g} = 4.44 g cm$^{-1}$, and [Fe/H] = 0.0 dex for the Sun as the reference star. 
	
	\begin{splitdeluxetable*}{cccccccccBcccccccBccccccc}
		\label{tab:EW}
		\tablecaption{Atomic Spectra line and Equivalent Width}
		\tabletypesize{\scriptsize}
		\colnumbers
		\tablehead{
			\colhead{$\lambda^1$} & \colhead{Ion$^1$} & \colhead{$X^1_{exc}$} & \colhead{log(gf)$^1$} & \colhead{$C_6^1$} & \colhead{Solar$^2$} & \colhead{TOI-1036$^2$} & \colhead{TOI-1055$^2$} & \colhead{TOI-1076$^2$} & \colhead{TOI-1097$^2$} & \colhead{TOI-1117$^2$} & \colhead{TOI-1203$^2$} & \colhead{TOI-215$^2$} & \colhead{TOI-2011$^2$} & \colhead{TOI-2426$^2$} & \colhead{TOI-3342$^2$} & \colhead{TOI-4628$^2$} & \colhead{TOI-4914$^2$} & \colhead{TOI-5005$^2$} & \colhead{TOI-5795$^2$} & \colhead{TOI-744$^2$} & \colhead{TOI-755$^2$} & \colhead{TOI-818$^2$} \\
			\colhead{\AA} & \colhead{} & \colhead{eV} & \colhead{} & \colhead{} & \colhead{EW} & \colhead{EW} & \colhead{EW} & \colhead{EW} & \colhead{EW} & \colhead{EW} & \colhead{EW} & \colhead{EW} & \colhead{EW} & \colhead{EW} & \colhead{EW} & \colhead{EW} & \colhead{EW} & \colhead{EW} & \colhead{EW} & \colhead{EW} & \colhead{EW} & \colhead{EW} 
		} 
		\startdata
		\hline
		5044.211 & 26.0	& 2.8512	& -2.058	& 2.71E-31	& 80.0	& 91.8	& 71.9 &	87.9	& 78.8	& 93	& 58.4	& 92.6	& 65.8	& 82.5	& 87.9	& 80.4	& 60.4	& 88.2	& 61.5	& 79.9	& 69.6	& 91.8 \\
		5054.642	& 26.0	& 3.64	& -1.921 &	4.68E-32	& 45.3	& 53.8	& 46.2	& 55.1	& 39.4	& 48	& 28.1	& 57.3	& 33.6	& 43.7	& 45.9	& 46	& 57.2	& 57.3	& 35.4	& 44.2	& 47.8	& 42.7 \\
		...	& ...	& ...	& ...	& ...	& ...	& ...	& ...	& ...	& ... & ...	& ... & ...	& ...	& ...	& ... & ...	& ...	& ...	& ... & ...	& ...	& ... \\          
		\hline
		\enddata
		\tablecomments{1. Columns 1-5 include the wavelength in angstroms (\AA), Ion numeric identifier, excitation potential, Log of the oscillator strength times the statistical weight, and damping constant for the atomic lines. The integer part of the Ion numeric identifier corresponds to the atom number, while the fractional part indicates the excitation state: ``.0" denotes the neutral line, and ``.1" denotes the first excitation line. For example, ``26.0" refers to the Fe I line, ``26.1" corresponds to the Fe II line, and ``6.0" indicates the C I line. \\
			2. Columns 6-23 are the equivalent width measurements for the Sun and the 17 targets. \\
			(This table is available in its entirety in machine-readable form online.)}
	\end{splitdeluxetable*}
	
	We measure the EWs for both the Fe I and Fe II lines in the object and solar spectra, and then derive $T_{\rm eff}$ by enforcing excitation balance for Fe I lines relative to the solar reference star. We compute log $g$ by enforcing ionization balance for Fe I and Fe II lines compared to the solar reference. The microturbulence is adjusted until the abundance variations show no correlation (p-value) with the reduced equivalent width (log EW/$\lambda$). We iterate the procedure until it converges to the outputs, which we adopt as the final stellar atmosphere parameters. We employ an error estimation method similar to that described in detail by \citet{2010ApJ...709..447E}. This approach takes into account the covariances among stellar parameters and the standard deviation of the differential abundances. We use the MARCS model atmosphere grid (\citealt{2008AA...486..951G}) for these calculations. The resulting $T_{\rm eff}$, log $g$, $V_t$, and their respective errors are detailed in Table \ref{tab:parameters}, while the final adopted [Fe/H] abundances are displayed in Table 4. We measure stellar atmospheric parameters to high precision, with errors of $T_{\rm eff}$ ranging from 10 to 46 K, averaging 26 K; errors of log $g$ are 0.03 to 0.08, averaging 0.06; errors of $V_t$ ranging from 0.05 to 0.11 km s$^{-1}$, averaging 0.07 km s$^{-1}$; and the typical errors of [Fe/H] ranging from 0.01 to 0.02 dex, except for TOI-215, which has an [Fe/H] error of 0.059 dex.
	
	\subsection{Abundance Analysis}
	
	We follow the methodology described in detail in \citet{2020AJ....159..220S, 2022MNRAS.513.5387S} to perform a line-by-line differential abundance analysis. In short, for chemical abundances of elements other than iron, we use the measured equivalent widths (EWs) and the final stellar atmospheric parameters to calculate the 1D-LTE abundances for each line in each star. For high precision abundance analysis, we conduct a differential abundance analysis strictly relative to the Sun. The abundances for the Sun and the 17 targets have been computed using the \textit{abfind} driver in MOOG. We present the abundance for each line of every star in Table \ref{tab:line_abund}. For each line for each star, we subtract the solar A(X) from the stellar A(X) to derive the [X/H] for that specific line for that star. We then calculate the mean [X/H] and standard deviation ($\sigma$) in the linear space by using all the lines for element X. By removing the outlying abundances that deviate more than twice the standard deviation from the mean [X/H], we compute the refined mean [X/H], the standard deviation ($\sigma$), and the standard deviation of the mean ($\sigma_{\mu}$ = $\sigma$/$\sqrt{N}$). The standard deviation of the mean is calculated by dividing the standard deviation by the square root of the number of lines used (N).
	
	\begin{splitdeluxetable*}{cccccccccBcccccccBccccccc}
		\label{tab:line_abund}
		\tablecaption{Abundance for each line of every star}
		\tabletypesize{\scriptsize}
		\colnumbers
		\tablehead{
			\colhead{$\lambda^1$} & \colhead{Ion$^1$} & \colhead{$X^1_{exc}$} & \colhead{log(gf)$^1$} & \colhead{$C_6^1$} & \colhead{Solar$^2$} & \colhead{TOI-1036$^2$} & \colhead{TOI-1055$^2$} & \colhead{TOI-1076$^2$} & \colhead{TOI-1097$^2$} & \colhead{TOI-1117$^2$} & \colhead{TOI-1203$^2$} & \colhead{TOI-215$^2$} & \colhead{TOI-2011$^2$} & \colhead{TOI-2426$^2$} & \colhead{TOI-3342$^2$} & \colhead{TOI-4628$^2$} & \colhead{TOI-4914$^2$} & \colhead{TOI-5005$^2$} & \colhead{TOI-5795$^2$} & \colhead{TOI-744$^2$} & \colhead{TOI-755$^2$} & \colhead{TOI-818$^2$} \\
			\colhead{\AA} & \colhead{} & \colhead{eV} & \colhead{} & \colhead{} & \colhead{A(X)} & \colhead{A(X)} & \colhead{A(X)} & \colhead{A(X)} & \colhead{A(X)} & \colhead{A(X)} & \colhead{A(X)} & \colhead{A(X)} & \colhead{A(X)} & \colhead{A(X)} & \colhead{A(X)} & \colhead{A(X)} & \colhead{A(X)} & \colhead{A(X)} & \colhead{A(X)} & \colhead{A(X)} & \colhead{A(X)} & \colhead{A(X)} 
		} 
		\startdata
		\hline
		5044.211	& 26.0	& 2.8512	& -2.058	& 2.71E-31	& 7.545	& 7.842	& 7.436	& 7.774	& 7.542	& 7.692	& 7.116	& 7.794	& 7.252	& 7.566	& 7.793	& 7.628	& 7.324	& 7.762	& 7.221	& 7.516	& 7.404	& 7.399 \\
		5054.642	& 26.0	& 3.64	& -1.921	& 4.68E-32	& 7.525	& 7.772	& 7.565	& 7.819	& 7.458	& 7.57	& 7.099	& 7.849	& 7.26	& 7.467	& 7.649	& 7.602	& 7.951	& 7.878	& 7.297	& 7.483	& 7.626	& 7.618 \\
		...	& ...	& ...	& ...	& ...	& ...	& ...	& ...	& ...	& ... & ...	& ... & ...	& ...	& ...	& ... & ...	& ...	& ...	& ... & ...	& ...	& ... \\          
		\hline
		\enddata
		\tablecomments{1. Columns 1-5 are the same as in Table \ref{tab:EW}. \\
			2. Columns 6-23 are the abundance measurements (A(X) in dex) for the Sun and the 17 targets. \\
			(This table is available in its entirety in machine-readable form online.)}
	\end{splitdeluxetable*}
	
	For elements with only one line, such as K and Sr, the 1$\sigma_{\mu}$ error corresponds to the 1$\sigma$ EW near that line. The 1$\sigma$ EW is calculated based on the S/N near that line, full-width at half-maximum (FWHM), and pixel scale (\citealt{2023ApJ...952...71S}):
	
	\begin{equation} \label{eqn1} 
		1 \sigma\ EW =1 \times 1.503 \times \frac{\sqrt{FWHM \times pixel\ scale }}{S/N}
	\end{equation}
	
	Apart from the measurement errors, we also consider systematic errors propagated from uncertainties in $T_{\rm eff}$, log $g$, and $V_t$. For example, we first adjust the $T_{\rm eff}$ values by their uncertainties to observe the resulting changes in $\sigma_{\mu}$. Similarly, we increase and decrease the values of log $g$ and $V_t$ to determine their respective impacts on $\sigma_{\mu}$. Next, we combine all changes in $\sigma_{\mu}$ from the variations in $T_{\rm eff}$, log $g$, and $V_t$ in quadrature, which represent the systematic errors. Finally, we add the systematic errors propagated from the stellar atmospheric parameters to the measured $\sigma_{\mu}$ in quadrature to obtain the final combined uncertainty. The final reported uncertainty on the elemental abundance shown in Table 4 is the quadrature sum of the two errors above. Furthermore, for Oxygen (O), we apply non-LTE (NLTE) corrections to the triplet around 7777 \AA\ to derive the final [O/H]. These NLTE corrections for O are applied to all stars using the grid provided by \citet{2007AA...465..271R}.
	
	\begin{figure}[!htbp]
		\centering
		\includegraphics[width=0.5\textwidth]{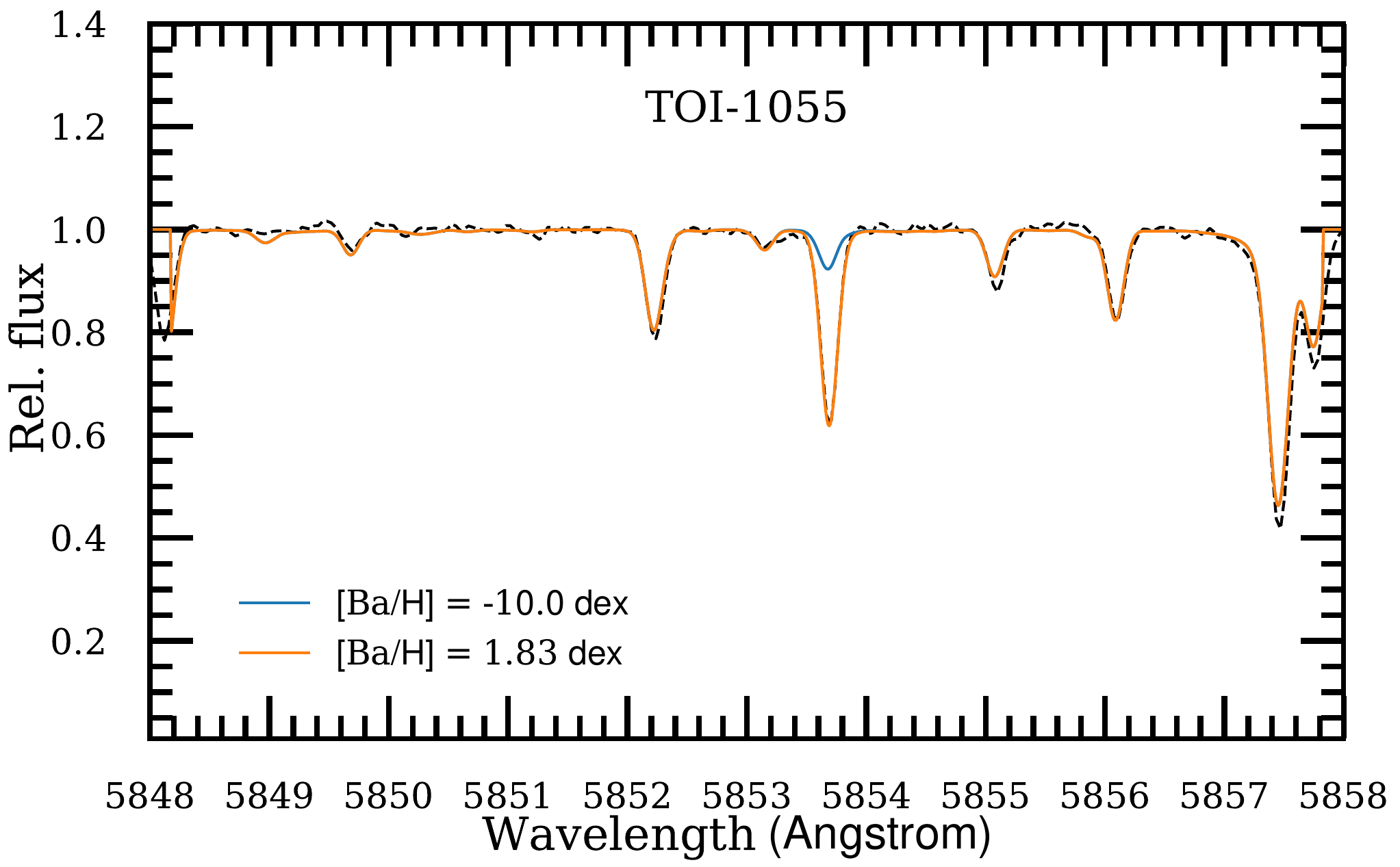}
		\caption{The observed and synthetic spectra for the Ba II 5853.67 \AA\ line for star TOI-1055. We retrieve the linelist around 10 \AA\ of the line using the linemake code (\citealt{2021RNAAS...5...92P}), and use the {\it synth} driver in MOOG to create synthetic spectra. The background dashed black line shows the observed spectra for TOI-1055 in this region. The best-fitted Ba abundance is shown as an orange line, while the no Ba case (-10.0 dex) is shown as a blue line for comparison.}
		\label{fig:synthesis}
	\end{figure}
	
	For elements with atomic numbers greater than 30, specifically Sr, Y, Ba, and Eu, we determine their abundances by creating synthetic spectra using the {\it synth} driver in MOOG for each element line. We begin with the linelist from Table 1 of \citet{2014ApJ...791...14M}. In the {\it synth} task in MOOG, we generate the linelist using the linemake code\footnote{\url{https://github.com/vmplacco/linemake/tree/master}} (\citealt{2021RNAAS...5...92P}), including all atomic and molecular lines within 10 \AA\ of the elemental line to aid in evaluating the continuum and fitting other lines. We perform synthesis on the Sr I 4607.338 line to derive abundances for Sr. For Y, we adopt the four Y II lines, including Y II 4883.685, Y II 4900.11, Y II 5087.42, Y II 5200.413, while excluding the Y II 4854.867 line due to its weak absorption feature and similarities to the non-detection threshold. For Ba, we synthesize the Ba II 5853.67, Ba II 6141.71, Ba II 6496.9 lines. In Figure \ref{fig:synthesis}, we show a sample synthesis of the Ba II 5853.67 \AA\ line for star TOI-1055. For Eu, we exclude the Eu II 3907.11 line due to severe blending issues. The Eu II 6645.1 line are weak for some of the stars, making confident detection challenging, so we only use abundance from cases where the detections are significant through visual inspection. The Eu II 4129.72 \AA\ line is good for the synthesis throughout. For easy reference, we present the final synthesized abundance for each line of the four elements in Table \ref{tab:line_abund}. We follow the same method described earlier to compute the averaged [X/H] and error. The average [X/H] and the combined uncertainties are shown in Table 4. In summary, we measure the differential [X/H] abundance for 23 elements, ranging from C to Eu, with average uncertainties range from 0.01 for elements such as Fe and Si, to 0.08 for elements like Sr, Y, and Eu.
	
	\begin{longrotatetable}
		\begin{deluxetable*}{ccccccccccccccccccc}
			\renewcommand\thetable{4}
			\tablecaption{Differential Abundances of 17 stars Relative to the Sun and their Uncertainties}
			\tabletypesize{\small}
			\tablehead{
				\colhead{Species$^1$} & \colhead{Atomic \#$^1$} & \colhead{50 \% $T_c^2$} & \multicolumn2c{TOI-1036} & \multicolumn2c{TOI-1055} & \multicolumn2c{TOI-1076} & \multicolumn2c{TOI-1097} & \multicolumn2c{TOI-1117} & \multicolumn2c{TOI-1203} & \multicolumn2c{TOI-215} & \multicolumn2c{TOI-2011} \\
				\colhead{} & \colhead{} & \colhead{K} &
				\colhead{[X/H]$^3$} & \colhead{err$^3$} & \colhead{[X/H]} & \colhead{err} & \colhead{[X/H]} & \colhead{err} & \colhead{[X/H]} & \colhead{err} & \colhead{[X/H]} & \colhead{err} & \colhead{[X/H]} & \colhead{err} & \colhead{[X/H]} & \colhead{err} & \colhead{[X/H]} & \colhead{err}
			} 
			\startdata
			\hline
			Fe	& 26 & 1334	& 0.204	& 0.012	& -0.003 & 0.012 & 0.378 & 0.018 & -0.010 & 0.015 &  0.136	& 0.020	& -0.399	& 0.015	 & 0.193	& 0.059	& -0.317	& 0.010 \\
			C	&  6 &   40	& 0.126	& 0.045	& -0.088 & 0.032 & 0.240 & 0.054 & -0.070 & 0.048 & -0.073	& 0.049	& -0.244	& 0.050	 & -0.185	& 0.065	& -0.229	& 0.028 \\
			O	&  8 &  180	& 0.059	& 0.023	& -0.020 & 0.025 & 0.234 & 0.029 & -0.065 & 0.064 &  0.043	& 0.041	& -0.120	& 0.049	 & 0.052	& 0.047	& -0.117	& 0.033 \\
			Na	& 11 &  958	& 0.141	& 0.051	& -0.060 & 0.023 & 0.408 & 0.057 & -0.037 & 0.051 &  0.172	& 0.033	& -0.293	& 0.051	 & 0.130	& 0.054	& -0.283	& 0.023 \\
			Mg	& 12 & 1336	& 0.154	& 0.039	& -0.023 & 0.027 & 0.327 & 0.082 & -0.036 & 0.048 &  0.162	& 0.040	& -0.147	& 0.029	 & 0.159	& 0.046	& -0.122	& 0.030 \\
			Al	& 13 & 1653	& 0.118	& 0.029	& -0.035 & 0.023 & 0.262 & 0.031 & -0.024 & 0.046 &  0.120	& 0.041	& -0.229	& 0.026	 & 0.067	& 0.038	& -0.178	& 0.033 \\
			Si	& 14 & 1310	& 0.162	& 0.011	& -0.009 & 0.008 & 0.354 & 0.010 & -0.007 & 0.012 &  0.128	& 0.017	& -0.272	& 0.011	 & 0.125	& 0.019	& -0.203	& 0.007 \\
			S	& 16 &  664	& 0.161	& 0.043	& -0.051 & 0.049 & 0.218 & 0.039 &  0.037 & 0.073 &  0.060	& 0.028	& -0.302	& 0.052	 & 0.003	& 0.101	& -0.349	& 0.043 \\
			K	& 19 & 1006	& 0.128	& 0.029	&  0.012 & 0.028 & 0.229 & 0.038 &  0.071 & 0.047 & -0.021	& 0.029	& -0.096	& 0.041	 & 0.061	& 0.047	&  0.005	& 0.030 \\
			Ca	& 20 & 1517	& 0.175	& 0.025	&  0.012 & 0.011 & 0.349 & 0.034 &  0.061 & 0.036 &  0.140	& 0.031	& -0.220	& 0.028	 & 0.163	& 0.042	& -0.187	& 0.019 \\
			Sc	& 21 & 1659	& 0.209	& 0.027	& -0.016 & 0.018 & 0.394 & 0.047 &  0.035 & 0.039 &  0.105	& 0.026	& -0.281	& 0.027	 & 0.179	& 0.025	& -0.194	& 0.022 \\
			Ti	& 22 & 1582	& 0.237	& 0.019	& -0.026 & 0.014 & 0.413 & 0.028 &  0.008 & 0.022 &  0.172	& 0.024	& -0.175	& 0.021	 & 0.202	& 0.025	& -0.105	& 0.017 \\
			V	& 23 & 1429	& 0.221	& 0.035	& -0.038 & 0.029 & 0.503 & 0.058 & -0.041 & 0.048 &  0.281	& 0.046	& -0.298	& 0.034	 & 0.241	& 0.054	& -0.168	& 0.027 \\
			Cr	& 24 & 1296	& 0.165	& 0.020	& -0.026 & 0.016 & 0.400 & 0.024 & -0.017 & 0.032 &  0.146	& 0.024	& -0.411	& 0.023	 & 0.226	& 0.028	& -0.337	& 0.016 \\
			Mn	& 25 & 1158	& 0.192	& 0.035	& -0.095 & 0.027 & 0.479 & 0.062 & -0.099 & 0.049 &  0.196	& 0.035	& -0.694	& 0.065	 & 0.219	& 0.018	& -0.566	& 0.034 \\
			Co	& 27 & 1352	& 0.196	& 0.036	& -0.027 & 0.020 & 0.414 & 0.027 & -0.056 & 0.052 &  0.170	& 0.026	& -0.272	& 0.041	 & 0.248	& 0.050	& -0.230	& 0.024 \\
			Ni	& 28 & 1353	& 0.182	& 0.019	& -0.040 & 0.018 & 0.394 & 0.021 & -0.038 & 0.027 &  0.131	& 0.016	& -0.396	& 0.015	 & 0.230	& 0.029	& -0.296	& 0.015 \\
			Cu	& 29 & 1037	& 0.117	& 0.061	& -0.094 & 0.041 & 0.299 & 0.063 & -0.029 & 0.028 &  0.125	& 0.064	& -0.364	& 0.082	 & 0.265	& 0.116	& -0.351	& 0.017 \\
			Zn	& 30 &  726	& 0.225	& 0.056	&  0.013 & 0.027 & 0.437 & 0.047 & -0.136 & 0.042 &  0.158	& 0.035	& -0.219	& 0.038	 & 0.231	& 0.029	& -0.174	& 0.033 \\
			Sr	& 38 & 1464	& 0.200	& 0.074	&  0.050 & 0.056 & 0.460 & 0.089 &  0.060 & 0.063 &  0.190	& 0.070	& -0.510	& 0.055	 & 0.200	& 0.085	& -0.380	& 0.041 \\
			Y	& 39 & 1659	& 0.043	& 0.042	& -0.001 & 0.026 & 0.351 & 0.087 &  0.063 & 0.040 &  0.005	& 0.051	& -0.575	& 0.052	 & 0.126	& 0.030	& -0.424	& 0.056 \\
			Ba	& 56 & 1455	& 0.191	& 0.031	&  0.067 & 0.024 & 0.394 & 0.024 &  0.194 & 0.078 &  0.054	& 0.028	& -0.608	& 0.057	 & 0.273	& 0.037	& -0.494	& 0.034 \\
			Eu	& 63 & 1356	& 0.215	& 0.031	& -0.005 & 0.025 & 0.290 & 0.084 &  0.090 & 0.077 &  0.000	& 0.080	&  0.051	& 0.036	 & 0.140	& 0.082	& -0.004	& 0.028 \\
			\hline
			\enddata
			\vspace{-10mm}
		\end{deluxetable*}
	\end{longrotatetable}
	
	\begin{longrotatetable}
		\begin{deluxetable*}{ccccccccccccccccccc}
			\renewcommand\thetable{4}
			\tablecaption{{\it (Continued)}}
			\tabletypesize{\small}
			\tablehead{
				\colhead{Species$^1$} & \multicolumn2c{TOI-2426} & \multicolumn2c{TOI-3342} & \multicolumn2c{TOI-4628} & \multicolumn2c{TOI-4914} & \multicolumn2c{TOI-5005} & \multicolumn2c{TOI-5795} & \multicolumn2c{TOI-744} & \multicolumn2c{TOI-755} & \multicolumn2c{TOI-818} \\
				\colhead{} & \colhead{[X/H]$^3$} & \colhead{err$^3$} & \colhead{[X/H]} & \colhead{err} & \colhead{[X/H]} & \colhead{err} & \colhead{[X/H]} & \colhead{err} & \colhead{[X/H]} & \colhead{err} & \colhead{[X/H]} & \colhead{err} & \colhead{[X/H]} & \colhead{err} & \colhead{[X/H]} & \colhead{err} & \colhead{[X/H]} & \colhead{err} 
			} 
			\startdata
			\hline
			Fe	& -0.025 & 0.007 &  0.132 & 0.016 &  0.120 & 0.019 & -0.015	& 0.021	&  0.215 &	0.015	& -0.199 & 0.014 & -0.036	& 0.009	&  0.094	& 0.015	&  0.086 &	0.013 \\
			C	& -0.258 & 0.040 & -0.143 & 0.027 & -0.079 & 0.056 & -0.256	& 0.072	& -0.287 &	0.057	& -0.151 & 0.050 & -0.140	& 0.035	& -0.131	& 0.029	& -0.141 &	0.029 \\
			O	& -0.129 & 0.023 & -0.009 & 0.026 & -0.057 & 0.064 & -0.215	& 0.033	& -0.003 &	0.029	& -0.080 & 0.041 & -0.061	& 0.024	& -0.049	& 0.032	&  0.025 &	0.033 \\
			Na	& -0.160 & 0.029 & -0.215 & 0.037 &  0.017 & 0.018 &  0.159	& 0.167	& -0.063 &	0.047	& -0.386 & 0.010 &  0.295	& 0.147	&  0.056	& 0.030	&  0.058 &	0.029 \\
			Mg	& -0.046 & 0.012 &  0.022 & 0.024 &  0.083 & 0.047 & -0.120	& 0.044	& -0.003 &	0.110	& -0.057 & 0.036 &  0.057	& 0.025	&  0.041	& 0.087	&  0.040 &	0.114 \\
			Al	&  0.001 & 0.015 &  0.065 & 0.023 &  0.048 & 0.034 & -0.215	& 0.030	& -0.056 &	0.047	& -0.089 & 0.034 & -0.087	& 0.013	& -0.033	& 0.039	& -0.029 &	0.041 \\
			Si	& -0.034 & 0.009 &  0.098 & 0.015 &  0.077 & 0.014 & -0.128	& 0.009	&  0.030 &	0.019	& -0.197 & 0.016 & -0.043	& 0.023	&  0.066	& 0.026	&  0.061 &	0.026 \\
			S	& -0.111 & 0.039 & -0.075 & 0.038 & -0.069 & 0.074 & -0.297	& 0.059	&  0.060 &	0.071	& -0.369 & 0.054 &  0.061	& 0.121	&  0.040	& 0.022	&  0.029 &	0.017 \\
			K	& -0.051 & 0.020 &  0.187 & 0.039 &  0.062 & 0.055 & -0.029	& 0.053	&  0.043 &	0.030	& -0.077 & 0.039 & -0.031	& 0.026	&  0.061	& 0.032	&  0.070 &	0.029 \\
			Ca	&  0.016 & 0.014 &  0.182 & 0.029 &  0.107 & 0.037 & -0.074	& 0.039	&  0.119 &	0.029	& -0.091 & 0.029 &  0.050	& 0.022	&  0.103	& 0.028	&  0.091 &	0.016 \\
			Sc	& -0.065 & 0.032 &  0.159 & 0.034 &  0.053 & 0.039 & -0.040	& 0.037	&  0.101 &	0.066	& -0.061 & 0.028 &  0.042	& 0.041	&  0.066	& 0.045	&  0.051 &	0.042 \\
			Ti	&  0.012 & 0.016 &  0.133 & 0.025 &  0.117 & 0.034 & -0.038	& 0.025	&  0.158 &	0.020	& -0.020 & 0.023 &  0.055	& 0.020	&  0.107	& 0.022	&  0.101 &	0.028 \\
			V	&  0.021 & 0.021 &  0.189 & 0.035 &  0.181 & 0.045 & -0.035	& 0.037	&  0.110 &	0.020	& -0.191 & 0.035 &  0.113	& 0.029	&  0.105	& 0.050	&  0.104 &	0.047 \\
			Cr	& -0.033 & 0.013 &  0.179 & 0.023 &  0.104 & 0.033 & -0.109	& 0.025	&  0.143 &	0.022	& -0.317 & 0.021 &  0.010	& 0.016	&  0.116	& 0.025	&  0.107 &	0.023 \\
			Mn	& -0.019 & 0.007 & -0.069 & 0.049 &  0.134 & 0.049 & -0.227	& 0.045	&  0.072 &	0.039	& -0.425 & 0.105 &  0.007	& 0.024	&  0.194	& 0.012	&  0.184 &	0.010 \\
			Co	& -0.035 & 0.017 & -0.049 & 0.040 &  0.110 & 0.038 & -0.045	& 0.041	&  0.048 &	0.031	& -0.226 & 0.034 &  0.072	& 0.022	&  0.107	& 0.018	&  0.117 &	0.019 \\
			Ni	& -0.041 & 0.013 &  0.101 & 0.022 &  0.098 & 0.026 & -0.097	& 0.030	&  0.091 &	0.015	& -0.291 & 0.026 &  0.037	& 0.020	&  0.138	& 0.022	&  0.132 &	0.020 \\
			Cu	&  0.017 & 0.061 & -0.001 & 0.015 &  0.086 & 0.035 & -0.146	& 0.029	& -0.045 &	0.031	& -0.159 & 0.150 &  0.035	& 0.036	&  0.131	& 0.045	&  0.123 &	0.045 \\
			Zn	& -0.054 & 0.017 & -0.037 & 0.032 &  0.093 & 0.021 & -0.162	& 0.038	&  0.084 &	0.056	& -0.126 & 0.026 &  0.007	& 0.019	&  0.156	& 0.043	&  0.138 &	0.072 \\
			Sr	& -0.150 & 0.068 &  0.200 & 0.067 &  0.150 & 0.092 & -0.050	& 0.077	&  0.380 &	0.079	& -0.390 & 0.076 &  0.060	& 0.064	&  0.260	& 0.072	&  0.190 &	0.077 \\
			Y	& -0.160 & 0.030 &  0.338 & 0.035 &  0.198 & 0.053 & -0.052	& 0.050	&  0.325 &	0.053	& -0.305 & 0.038 & -0.023	& 0.031	&  0.123	& 0.047	&  0.095 &	0.061 \\
			Ba	& -0.099 & 0.038 &  0.418 & 0.044 &  0.119 & 0.045 &  0.057	& 0.038	&  0.484 &	0.025	& -0.272 & 0.043 & -0.046	& 0.024	&  0.219	& 0.036	&  0.170 &	0.067 \\
			Eu	&  0.095 & 0.019 &  0.210 & 0.074 &  0.086 & 0.044 &  0.040	& 0.040	&  0.236 &	0.042	&  0.026 & 0.042 &  0.046	& 0.031	&  0.070	& 0.035	&  0.096 &	0.038 \\
			\hline
			\enddata
			\tablecomments{1. The name of the element and atom number, we have adopted the linelist from Table 1 of \citet{2014ApJ...791...14M} for atom number smaller than or equal to 30. For Sr, Y, Ba, Eu, we generated synthetic spectra for each of the lines.
				2. The 50\% equilibrium condensation temperature ($T_c$) for a Solar-System Composition Gas (\citealt{2003ApJ...591.1220L}). Here, 50\% means that half of the element is in the solid state, and the other half is in the gas state.
				3. Elemental abundance [X/H] and associated uncertainties (err). The uncertainty is combined from measurement errors ($\sigma_{\mu}$) and systematic errors propagated from 1-$\sigma$ uncertainty in $T_{\rm eff}$, log $g$, and $V_t$. The O abundance have been corrected for the NLTE effect given by \citet{2007AA...465..271R}.
			}
			\vspace{-10mm}
		\end{deluxetable*}
	\end{longrotatetable}
	
	\section{Stellar Isochrone age, mass, and radius}   \label{sec:age}
	
	We compare observed stellar properties with theoretical stellar isochrones to determine stellar age, mass, and radius. We use the q$^2$ package (\citealt{Ramirez2014}) in conjunction with the Yonsei-Yale isochrones (\citealt{2001ApJS..136..417Y}), which provide a systematic framework for comparing stellar properties to evolutionary models. The Yonsei-Yale isochrones offer a set of evolutionary tracks that include a range of metallicities and ages, and have been commonly employed in the past (e.g. \citealt{2020AJ....159..246S}).
	
	\citet{2012AA...543A..29M} suggest that incorporating spectroscopic log $g$ as an input parameter improves the precision of age estimates than using photometry when performing isochrone fitting. This is because spectroscopic measurements of log $g$ provide a direct constraint on the star's evolutionary state and structure, which can be more robust than using absolute $V$ magnitude derived from photometry. While absolute magnitude can be affected by interstellar extinction and distance uncertainties, the spectroscopic log $g$ is less susceptible to these observational biases, leading to more reliable age determinations.
	
	The $T_{\rm eff}$, log $g$, and [Fe/H] derived from high-resolution spectroscopy are set as the input parameters. The q$^2$ package performs a maximum likelihood estimation to find the most probable stellar age, radius, mass, and associated uncertainties. The final age, stellar radius, stellar mass, and their corresponding uncertainties are shown in Table \ref{tab:parameters}.
	
	Our measured stellar parameters may slightly alter the measured planetary parameters from the NASA Exoplanet Archive. For example, the stellar radius is used to compute the planetary radius. Our precisely determined stellar radius from high-resolution spectroscopy agree with those from the NASA Exoplanet Archive within 1$\sigma$ (10 stars), 2$\sigma$ (five stars, TOI-1055, TOI-215, TOI-2426, TOI-3342, TOI-818), and 3$\sigma$ (two stars, TOI-2011 \& TOI-744) of our derived errors.
	
	\section{The [X/Fe] trend and Galactic-chemical Evolution}   \label{sec:GCE}
	
	With the abundances of 22 elements and derived stellar ages, we now assess our results by comparing them to existing abundance measurements from large surveys. This comparison will help identify any anomalous abundance measurements in the stars and allow us to explore the potential impact of Galactic Chemical Evolution (GCE). Given the range of -0.4 to 0.4 dex in [Fe/H] for our stars, we adopt [X/Fe] instead of [X/H], providing an assessment of the abundance of element X relative to Fe in a star. We analyze the relationship between [X/Fe] and both the overall metallicity ([Fe/H]) and age. To explore the influence of planet formation on shifts in [X/Fe] and determine if planet formation causes statistically significant differences in abundance measurements, we compare our abundance data with that of solar-like stars from large surveys that have similar stellar parameter ranges.
	
	\subsection{The [X/Fe] -- [Fe/H] trend}
	
\begin{figure*}   
		\centering
		\includegraphics[width=1.0\textwidth, trim = {0 2cm 0 5cm}]{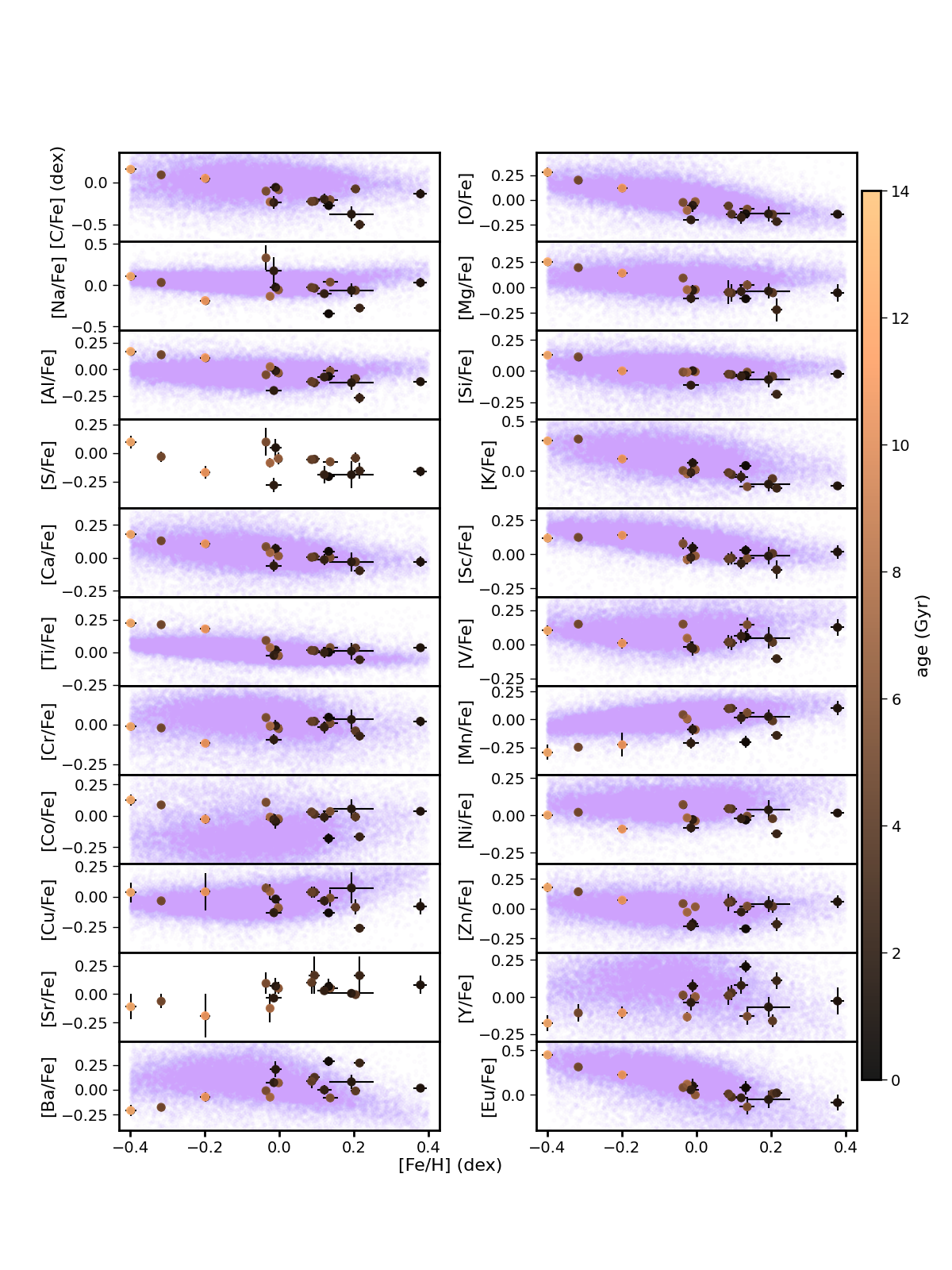}
		\caption{The [X/Fe] versus [Fe/H] trend for the 22 elements, color-coded based on their isochrone age. We compare our data to the background light purple points, which represent abundances from the GALAH DR2 survey (\citealt{2018MNRAS.478.4513B}). Since the GALAH DR2 survey does not report abundances for S and Sr, these elements are not shown. The background purple points are restricted to stars with $\Delta T_{\rm eff}$ $<$ 200 K, $\Delta \log g$ $<$ 0.2 dex, and $\Delta$[Fe/H] $<$ 0.4 dex relative to the Sun, comparable to the parameter ranges of the 17 solar-like stars listed in Table \ref{tab:parameters}. Overall, our abundance trends align with those from the GALAH survey.}
		\label{fig: XFe_FeH}
	\end{figure*}

	In Figure \ref{fig: XFe_FeH}, we display the trends of [X/Fe] abundances versus [Fe/H] for the 22 elements, with each star color-coded by their isochrone age. We compare our abundances to those from the GALAH DR2 survey (\citealt{2018MNRAS.478.4513B}). We restrict the GALAH sample to stars with $\Delta T_{\rm eff}$ $<$ 200 K, $\Delta \log g$ $<$ 0.2 dex, and $\Delta$[Fe/H] $<$ 0.4 dex, similar to the parameter ranges of our 17 solar-like stars. Overall, our abundance trends are generally consistent with those from GALAH. Occasionally, a few metal-poor stars (with [Fe/H] $<$ 0.2 dex) and metal-rich stars (with [Fe/H] $>$ 0.2 dex) do not align with the majority of the GALAH data. Additionally, GALAH does not report abundances for sulfur (S) and strontium (Sr), so these elements are not included in our comparison.
	
\begin{figure*}  
		\centering			
		\includegraphics[width=1.0\textwidth, trim = {0 2cm 0 5cm}]{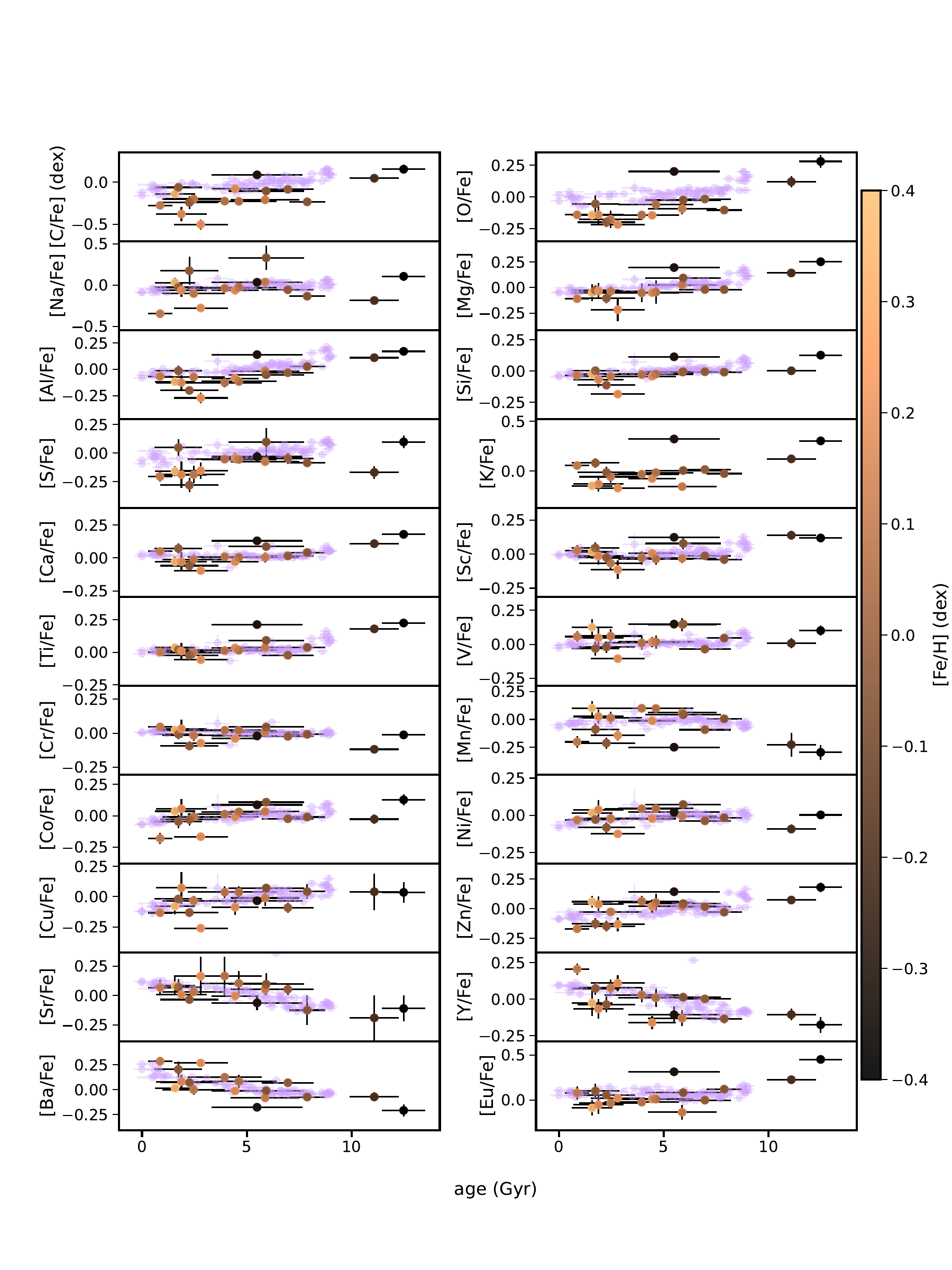}
		\caption{The [X/Fe] versus age trend for 22 elements, color-coded by their [Fe/H]. We compare our findings with the sample of 68 solar twins from \citet{Bedell2018}. The isochrone ages and abundances of the four neutron capture elements for this sample are detailed in \citet{2018MNRAS.474.2580S}. Since potassium abundance is not available for this sample, we exclude it from the plot. Additionally, the carbon and oxygen abundances in our sample are consistently lower than those of the solar twins from \citet{Bedell2018} by an average of 0.186 and 0.096 dex, respectively, using solar twins only.}
		\label{fig: XFe_age}
	\end{figure*}

	Figure \ref{fig: XFe_age} displays the [X/Fe] trends in relation to stellar isochrone age, color-coded based on their [Fe/H]. Lighter elements such as C, O, Na, and Mg show higher abundance levels in older stars when compared to the rest, while heavier elements like Sr, Y, and Ba show notably lower abundance. This imply that the older stars, formed early in the Galaxy, may not have sufficient time to produce these heavier elements through nucleosynthesis when they were formed. We compare our findings with the sample of 68 solar twins ($\Delta T_{\rm eff}$ $<$ 100 K, $\Delta log g$ $<$ 0.1 dex, and $\Delta$[Fe/H] $<$ 0.1 dex) from \citet{2018MNRAS.474.2580S} and \citet{Bedell2018}, where empirical Galactic Chemical Evolution (GCE) correlations have also been derived. Since potassium abundance is not available for this sample, we exclude it from the plot. We perform linear regression fits (emcee: 2013PASP..125..306F) for each element, considering the error bars of abundance and using solar twins only. The slopes, intercepts, and errors are shown in Table 5, along with the slopes from \citet{Bedell2018} for easy comparison.
	
	\begin{deluxetable*}{ccccc}[!htbp]
		\renewcommand\thetable{5}
		\tablecaption{The linear regression coefficients for the fits of [X/Fe] versus age for 22 elements}
		\tabletypesize{\scriptsize}
		\tablehead{
			Species & m$^1$ & b$^1$ & m$^2$ & b$^2$ \\
			& dex Gyr$^{-1}$ & dex & dex Gyr$^{-1}$ & dex
		} 
		\startdata
		\hline
		C &	0.0163$_{-0.0250}^{+0.0250}$	& -0.2693$_{-0.1511}^{+0.1511}$ & 0.0115 $\pm$ 0.0014	& −0.0836 $\pm$ 0.0089  \\
		O &	0.0089$_{-0.0178}^{+0.0178}$ & -0.1217$_{-0.1078}^{+0.1078}$ & 0.0088 $\pm$ 0.0014	& −0.0260 $\pm$ 0.0075  \\
		Na &	-0.0349$_{-0.0578}^{+0.0189}$ & 0.1743$_{-0.1190}^{+0.3786}$	& 0.0086 $\pm$ 0.0016	& −0.0614 $\pm$ 0.0089  \\
		Mg & -0.0254$_{-0.0300}^{+0.0184}$ & 0.1712$_{-0.1170}^{+0.1875}$	& 0.0099 $\pm$ 0.0009	& −0.0367 $\pm$ 0.0048  \\
		Al &	0.0393$_{-0.0036}^{+0.0036}$ & -0.2874$_{-0.0219}^{+0.0219}$	& 0.0139 $\pm$ 0.0010	& −0.0595 $\pm$ 0.0054  \\
		Si &	0.0042$_{-0.0019}^{+0.0019}$ & -0.0392$_{-0.0113}^{+0.0113}$	& 0.0063 $\pm$ 0.0006	& −0.0308 $\pm$ 0.0033  \\
		S &	-0.0047$_{-0.0256}^{+0.0256}$ & -0.0331$_{-0.1550}^{+0.1550}$	& 0.0098 $\pm$ 0.0015	& −0.0537 $\pm$ 0.0085  \\
		K &	0.0006$_{-0.0068}^{+0.0068}$ & -0.0157$_{-0.0410}^{+0.0410}$	& --& --  \\
		Ca &	0.0062$_{-0.0112}^{+0.0112}$ & -0.0081$_{-0.0675}^{+0.0675}$	& -0.0011 $\pm$ 0.0006	& 0.0217 $\pm$ 0.0032  \\
		Sc &	-0.0159$_{-0.0227}^{+0.0109}$ & 0.0917$_{-0.0650}^{+0.1387}$	& 0.0059 $\pm$ 0.0009	& -0.0263 $\pm$ 0.0052  \\
		Ti &	-0.0115$_{-0.0238}^{+0.0117}$ & 0.0981$_{-0.0686}^{+0.1440}$	& 0.0036 $\pm$ 0.0005	& -0.0024 $\pm$ 0.0032  \\
		V &	-0.0199$_{-0.0386}^{+0.0194}$ & 0.1634$_{-0.1137}^{+0.2336}$ & 0.0013 $\pm$ 0.0007	& -0.0023 $\pm$ 0.0037  \\
		Cr &	-0.0135$_{-0.0116}^{+0.0086}$ & 0.0941$_{-0.0545}^{+0.0723}$ & -0.0016 $\pm$ 0.0003	& 0.0095 $\pm$ 0.0019  \\
		Mn &	-0.0380$_{-0.0269}^{+0.0191}$ & 0.2587$_{-0.1161}^{+0.1598}$ & 0.0023 $\pm$ 0.0012	& -0.0312 $\pm$ 0.0063  \\
		Co &	-0.0184$_{-0.0258}^{+0.0146}$ & 0.1333$_{-0.0860}^{+0.1564}$ & 0.0074 $\pm$ 0.0011	& -0.0460 $\pm$ 0.0057  \\
		Ni &	-0.0225$_{-0.0167}^{+0.0126}$ & 0.1556$_{-0.0778}^{+0.1024}$ & 0.0071 $\pm$ 0.0009	& -0.0505 $\pm$ 0.0050  \\
		Cu &	-0.0259$_{-0.0303}^{+0.0189}$ & 0.1708$_{-0.1103}^{+0.1819}$ & 0.0149 $\pm$ 0.0017	& -0.0850 $\pm$ 0.0097  \\
		Zn &	-0.0266$_{-0.0140}^{+0.0133}$ & 0.1890$_{-0.0876}^{+0.0924}$ & 0.0102 $\pm$ 0.0014	& -0.0699 $\pm$ 0.0077  \\
		Sr &	-0.05917$_{-0.0480}^{+0.0383}$ & 0.4247$_{-0.2417}^{+0.2986}$ & -0.0251 $\pm$ 0.0030	& 0.1310 $\pm$ 0.0164  \\
		Y &	-0.0390$_{-0.0242}^{+0.0192}$ & 0.2185$_{-0.1205}^{+0.1510}$ & -0.0238 $\pm$ 0.0024	& 0.1135 $\pm$ 0.0130  \\
		Ba & -0.0414$_{-0.0301}^{+0.0230}$ & 0.2840$_{-0.1415}^{+0.1836}$ & -0.0317 $\pm$ 0.0018	& 0.1897 $\pm$ 0.0093  \\
		Eu & 0.0277$_{-0.0155}^{+0.0155}$ & -0.1252$_{-0.0934}^{+0.0934}$ & -0.0056 $\pm$ 0.0017	& 0.0908 $\pm$ 0.0093  \\
		\hline
		\enddata
		\tablecomments{1. Slope and intercept with errors for 22 elements, using emcee for linear regression fits. The linear fit is performed using solar twins only. \\
			2. Slope and intercept with errors for 21 elements (K not included) from \citet{Bedell2018}.}
	\end{deluxetable*}
	
	The carbon and oxygen abundances in our sample are consistently lower than those of the solar twins in \citet{Bedell2018} by an average of 0.186 and 0.096 dex, respectively, using solar twins only. Although we use the same line list for carbon and oxygen and apply similar NLTE corrections, the reason for the systematically low abundances remains unclear. Additionally, when comparing our results to the GALAH survey, we find that the carbon and oxygen abundances are consistent with GALAH. It is possible that our sample simply has lower carbon and oxygen abundances. Note that the slopes of [C/Fe] versus age and [O/Fe] versus age agree with those reported in \citet{Bedell2018} (shown by the light purple background) within 1$\sigma$ of our uncertainty as shown in Table 5. Consequently, we apply the same Galactic Chemical Evolution (GCE) correction as they did, given that they have more data points for the linear analysis, and this is appropriate we are only correcting for the trend effect. Aside from C and O, no other elements exhibit a noticeable offset compared to the solar twins in \citet{Bedell2018}.
	
	\subsection{Galactic Chemical Evolution}
	
	Based on Figure \ref{fig: XFe_age}, the subtle trends of [X/Fe] versus age suggest the influence of Galactic Chemical Evolution (GCE) effects. Previous works suggested that GCE may impact the condensation temperature ($T_c$) trends, and that the [X/Fe] ratio is associated with stellar age and distance from the Galactic center, as indicated by studies such as \cite{2014A&A...564L..15A, 2015A&A...579A..52N, 2016A&A...593A.125S, Bedell2018}. Therefore, we apply GCE corrections to the elemental abundances before looking into the link between $T_c$ trends and planet formation for solar twins.
	
	Our current sample includes 17 stars with varying metallicities around the Sun, and the trend of [X/Fe] with age is comparable to that observed in \cite{Bedell2018}. Given the relatively small size of our sample and the metallicity ranges it spans, along with the subtle nature of the Galactic Chemical Evolution (GCE) effect, we adopt the GCE relations derived by \cite{Bedell2018}, which determines the abundances of 30 elements for 79 Sun-like stars (68 solar twins) within 100 pc, achieving a precision of 2\%. It finds that stars with similar ages and metallicities nearly have identical abundance patterns. Their GCE correlation is derived specifically for solar twins, which satisfy the criteria of $\Delta T_{\rm eff} < 100$ K, $\Delta \log g < 0.1$ dex, and $\Delta [Fe/H] < 0.1$ dex from the Sun. Based on Tables \ref{tab:parameters} and 4, we identify five stars in our sample as solar twins: TOI-1055, TOI-2426, TOI-744, TOI-755, and TOI-818. Using the coefficients from Table 3 in \cite{Bedell2018} along with our isochrone ages, we apply GCE corrections to all elements in these five solar twins, except for potassium (K), which is not covered by \cite{Bedell2018}. For the GCE-corrected abundances, we have combined uncertainties propagated from the coefficients of the GCE correction.
	
	Ideally, we plan to develop our own GCE corrections while considering the effects of metallicity, as there may be systematic differences between studies. Nevertheless, given the size of our current sample and the metallicity ranges of the stars we are studying, we currently use GCE correction relations from existing literature. In upcoming observations for the PASTA survey, we will focus on including more solar twins/analogs. With a larger sample, we will be able to create our own GCE relations and compare them to existing studies.
	
	\section{Condensation Temperature Trend}   \label{sec:Tc}
	
	In this section, we analyze the differential abundance relative to the Sun versus the condensation temperature trend for each of the 17 stars. Specifically, we focus on the trend for the 5 solar twins after applying GCE corrections to their abundances. This analysis aims to determine whether the Sun shows unique abundance characteristics compared to the solar twins and to explore the potential impact of planet formation scenarios on these differences.
	
	\subsection{Derivation of the $T_c$ trend}
	
\begin{figure*}
		\centering			
		\includegraphics[width=0.9\textwidth]{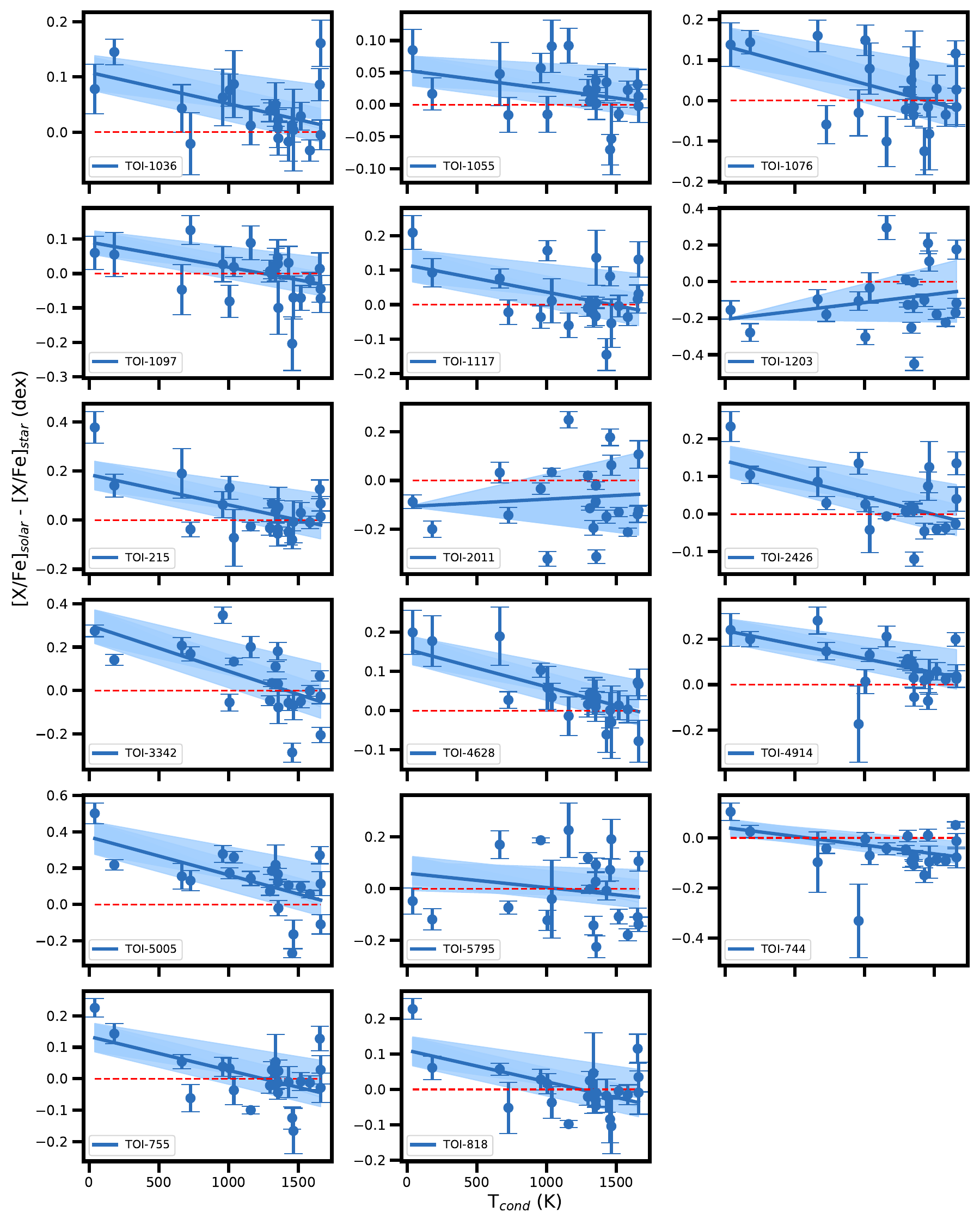}
		\caption{The trend of differential elemental abundance ([X/Fe]$_{\rm solar}$ - [X/Fe]$_{\rm star}$) versus $T_c$, with a linear fit to the blue symbols that include error bars, shown by a blue line. The background blue band displays the 1$\sigma$ confidence interval. The Sun is located at a value and slope of 0, shown by the red dashed horizontal line.}
		\label{fig:Tc_trend}
	\end{figure*}

	Following the determination of the differential abundances and their respective uncertainties, we proceed to investigate the differential abundance with $T_c$ trends. We adopt the 50\% equilibrium condensation temperature from \citet{2003ApJ...591.1220L}, which is computed based on solar system composition gas. We adopt a $T_c$ of 1300 K as the threshold for distinguishing between refractory and volatile elements; those above this temperature are considered refractory elements, while those below are considered volatile elements, as suggested by \citet{2001sse..book.....T, 2021SSRv..217...44L, 2020NatAs...4..314M}. As the metallicity of our stars span a range of metallicities around the Sun, we compute [X/Fe] of each element, and then compute the difference between solar abundance and that of the star ([X/Fe]$_{\rm solar}$ - [X/Fe]$_{\rm star}$, equiqvalent to $-$[X/Fe]). In Figure \ref{fig:Tc_trend}, we show [X/Fe]$_{\rm solar}$ - [X/Fe]$_{\rm star}$ versus $T_c$ for the 17 stars.
	
	\begin{figure*}[!htbp]
		\centering	\includegraphics[width=0.9\textwidth]{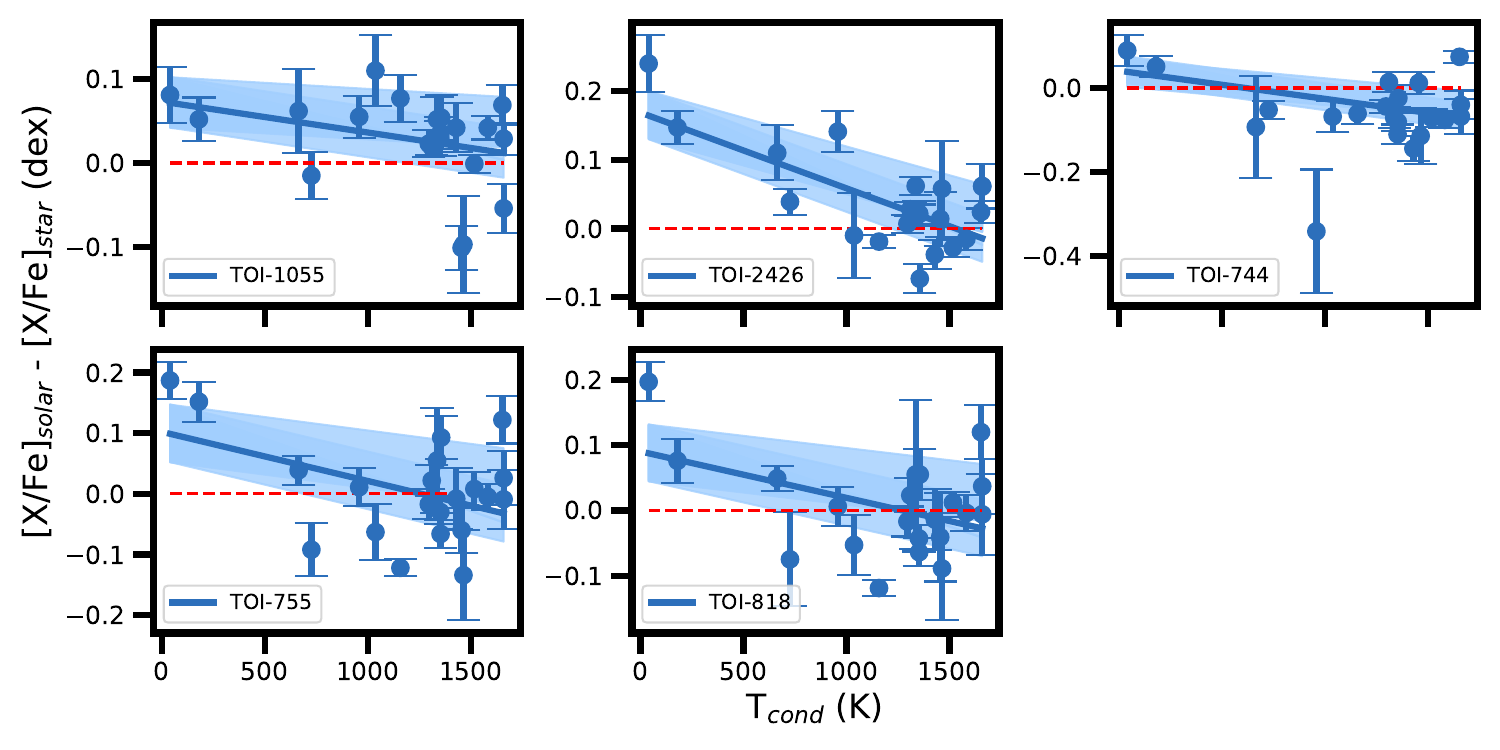}
		\caption{Trend of the GCE-corrected differential elemental abundance ([X/Fe]$_{\rm solar}$ - [X/Fe]$_{\rm star}$) versus $T_c$ for five solar twins. The blue symbols show the data points with associated error bars. A linear fit is applied to these data points, shown by the blue line. Error bars are incorporated into the linear fit. The shaded blue band in the background is the 1 $\sigma$ confidence interval. The Sun is located at a value and slope of 0, shown by the red dashed horizontal line.}
		\label{fig:Tc_GCE_trend}
	\end{figure*}
	
	We perform a linear regression fit to the differential abundance versus $T_c$ plotconsidering error bars. A similar method for linear regression fitting that accounts for error bars using emcee is discussed in \cite{2024AJ....167..167S}. In Figure \ref{fig:Tc_trend}, we plot the best-fit linear regression and the 1$\sigma$ confidence interval. For the five solar twins (TOI-1055, TOI-2426, TOI-744, TOI-755, and TOI-818), we apply GCE corrections to their abundances and perform a similar linear regression analysis, as shown in Figure \ref{fig:Tc_GCE_trend}. Since Potassium has not been corrected for GCE, we omit K from the linear regression analysis and focus on the remaining elements that have been corrected. After GCE correction, the $T_c$ trend slope becomes flatter than the original slope. In Figure \ref{fig:Tc_GCE_trend}, the $-$[X/Fe] vs. $T_c$ slopes of these solar twins are all negative, suggesting that the Sun is more depleted in refractory elements than volatile elements compared to solar twins.  In Table 6, we list the final $T_c$ trend slope and intercept of the linear fit for each of the stars.
	
	\begin{longrotatetable}
		\begin{deluxetable*}{ccccccccccccccc}
			\renewcommand\thetable{6}
			\tablecaption{T$_c$ Trend Slope and Intercept with Average Abundances of Volatile and Refractory Elements}
			\tabletypesize{\scriptsize}
			\tablehead{
				\colhead{name$^1$} & \colhead{slope$^1$} & \colhead{err$_l^1$} & \colhead{err$_u^1$}	& \colhead{intercept$^1$}	& \colhead{err$_l^1$} & \colhead{err$_u^1$} &
				\colhead{slope$_{GCE}^2$} & \colhead{err$_l^2$} & \colhead{err $_u^2$} & \colhead{intercept$_{GCE}^2$} & \colhead{err$_l^2$} & \colhead{err$_u^2$} & \colhead{$\overline{[X/H]}_{volatile}^3$} &\colhead{$\overline{[X/H]}_{refractory}^3$} 
			}
			\colnumbers
			\startdata
			\hline
			TOI-1036 & -5.70e-05 & 2.23e-05 & 2.43e-05    &  0.108  & 0.031  & 0.029       &  -- & -- & --            &   --  & --  & -- & -- & -- \\
			TOI-1055 & -2.87e-05 & 1.80e-05 & 1.72e-05    &  0.053  & 0.023  & 0.024       & -3.69e-05 & 2.36e-05 & 2.24e-05            &   0.073  & 0.030  & 0.031 & 0.06 $\pm$ 0.027 & 0.010 $\pm$ 0.015 \\
			TOI-1076 & -9.34e-05 & 3.64e-05 & 3.64e-05    &  0.135  & 0.046  & 0.046       & -- & -- & -- &   --  & -- & -- & -- & -- \\
			TOI-1097 & -7.33e-05 & 2.60e-05 & 2.58e-05    &  0.091  & 0.034  & 0.034       & -- & -- & -- &   --  & --  & -- & -- & -- \\
			TOI-1117 & -7.81e-05 & 3.67e-05 & 3.59e-05    &  0.115  & 0.046  & 0.047       & -- & -- & -- &   --  & -- & -- & -- & --\\
			TOI-1203 &  9.16e-05 & 1.01e-04 & 1.03e-04    & -0.206  & 0.000  & 0.000       &  -- & -- & -- &  --  & --  & -- & -- & -- \\
			TOI-215 & -1.24e-04 & 4.57e-05 & 4.35e-05    &  0.186  & 0.057  & 0.059       & -- & -- & -- & --  & -- & -- & -- & -- \\
			TOI-2011 &  3.03e-05 & 1.01e-04 & 1.03e-04    & -0.107  & 0.000  & 0.000       &  -- & -- & -- &  --  & --  & -- & -- & -- \\
			TOI-2426 & -9.51e-05 & 3.26e-05 & 3.25e-05    &  0.141  & 0.041  & 0.042       & -1.10e-04 & 2.70e-05 & 2.65e-05            &   0.169  & 0.034  & 0.035 & 0.093 $\pm$ 0.056 & 0.015 $\pm$ 0.016 \\
			TOI-3342 & -2.13e-04 & 6.10e-05 & 6.04e-05    &  0.301  & 0.076  & 0.077       & -- & -- & -- &   --  & -- & -- & -- & -- \\
			TOI-4628 & -9.54e-05 & 2.92e-05 & 2.74e-05    &  0.156  & 0.035  & 0.037       & -- & -- & -- &   --  & -- & -- & -- & -- \\
			TOI-4914 & -1.24e-04 & 4.11e-05 & 4.09e-05    &  0.238  & 0.053  & 0.053       & -- & -- & -- &   -- & --  & -- & -- & -- \\
			TOI-5005 & -2.10e-04 & 6.91e-05 & 6.89e-05    &  0.372  & 0.087  & 0.088       & -- & -- & -- &   -- & --  & -- & -- & -- \\
			TOI-5795 & -5.55e-05 & 5.22e-05 & 3.87e-05    &  0.060  & 0.042  & 0.067       & -- & -- & -- &   -- & -- & -- & -- & -- \\
			TOI-744 & -6.99e-05 & 2.71e-05 & 2.13e-05    &  0.041  & 0.027  & 0.035       & -6.69e-05 & 2.82e-05 & 2.19e-05            &   0.041  & 0.027  & 0.037 & -0.065 $\pm$ 0.052 & -0.053 $\pm$ 0.018 \\
			TOI-755 & -1.08e-04 & 3.56e-05 & 3.55e-05    &  0.134  & 0.044  & 0.045       & -8.10e-05 & 3.82e-05 & 3.68e-05            &   0.102  & 0.047  & 0.048 & 0.016 $\pm$ 0.062 & -0.000 $\pm$ 0.039 \\
			TOI-818 & -8.88e-05 & 3.19e-05 & 3.22e-05    &  0.110  & 0.040  & 0.040       & -7.10e-05 & 3.51e-05 & 3.37e-05            &   0.090  & 0.043  & 0.044 & 0.011 $\pm$ 0.057 & 0.002 $\pm$ 0.035 \\
			\hline
			\enddata
			\tablecomments{1. Columns 1-7 are the star name, the T$_c$ slope, 1$\sigma$ lower error, and 1$\sigma$ upper error, the intercept, 1$\sigma$ lower error, and 1$\sigma$ upper error from the linear regression.
				2. The same parameters as column 2-7, but GCE-corrected for solar twins using the relation from \cite{Bedell2018}.
				3. Average differential abundance ([X/Fe]$_{\rm solar}$ - [X/Fe]$_{\rm star}$) of volatile and refractory elements for the five solar twins after GCE correction. The errors are propagated from individual measurements.}
		\end{deluxetable*}
	\end{longrotatetable}
	
	\subsection{Direct Comparison Between the Sun and Solar Twins}
	
\begin{figure*}
		\centering
		\includegraphics[width=0.75\textwidth, trim={4cm 0 4cm 0}]{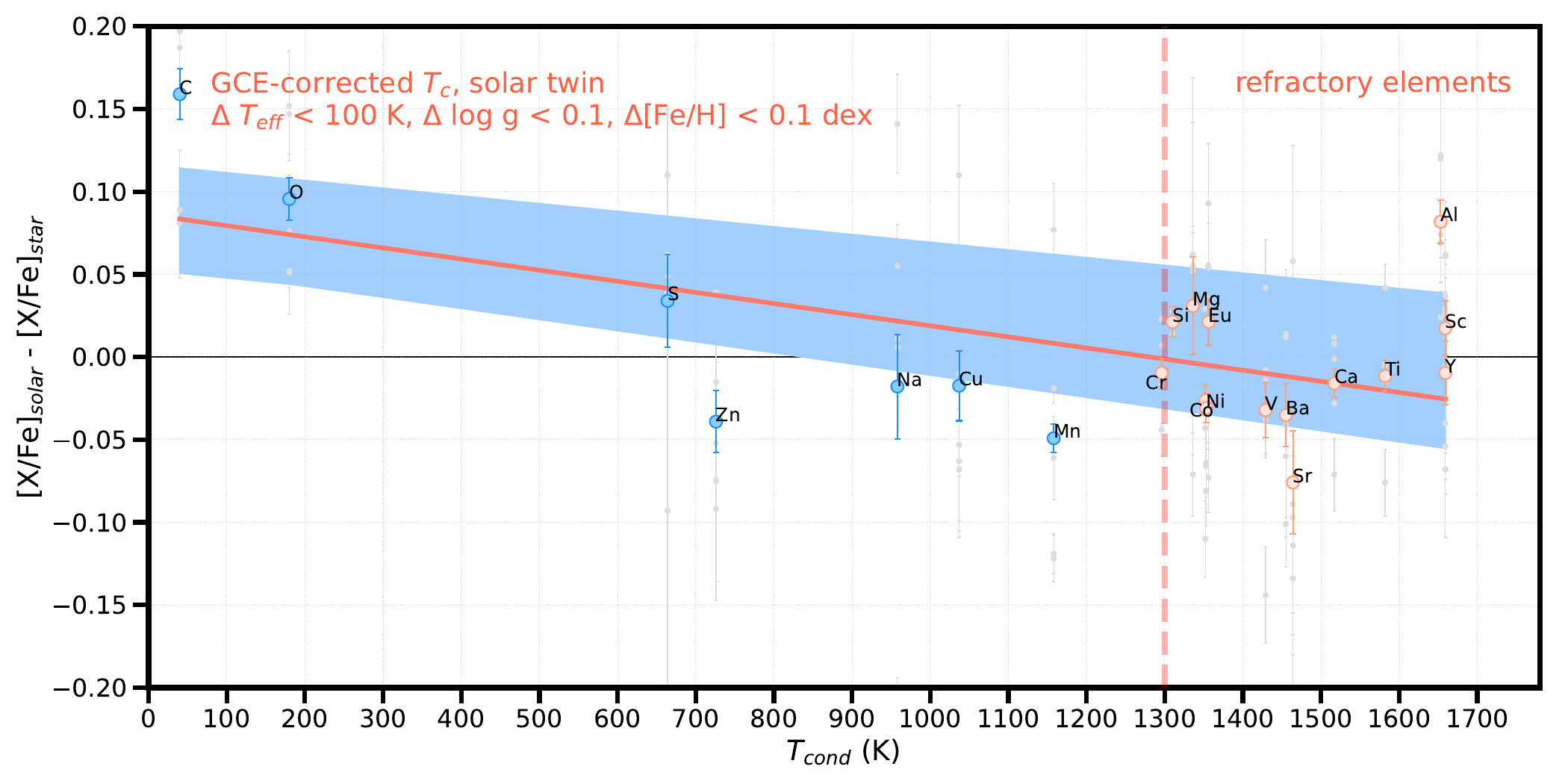}
		\caption{The GCE-corrected, differential elemental abundance relative to the Sun ([X/Fe]$_{\rm solar}$ - [X/Fe]$_{\rm star}$). The five solar twins with -0.1 $<$ [Fe/H] $<$ 0.1 dex, $\Delta\ T_{eff}$ $<$ 100 K, $\Delta$ log g $<$ 0.1 are TOI-1055, TOI-2426, TOI-744, TOI-755, and TOI-818, are shown as gray symbols in the background, while the colored symbols show the average elemental abundance for these stars. Refractory elements are shown in orange, volatile elements are shown in blue symbols. Elements with $T_c$ above 1300 K are considered refractory, with the threshold $T_c$ of 1300 K shown by a dashed orange line. We fit a linear regression to the colored symbols while considering the error bars. The orange line displays the best linear fit to all the colored points, with the shaded blue band indicating the 1-$\sigma$ confidence interval. The expression for the linear regression is -6.726e-05$_{-2.342e-05}^{+2.320e-05}$ $\times$ $T_c$ (K) + 0.086$_{-0.030}^{+0.030}$.}
		\label{fig:differential}
	\end{figure*}

	We compare the average differential elemental abundance for the five solar twins relative to the Sun in Figure \ref{fig:differential}, where the GCE-corrected abundances are used. In the figure, the gray points in the background show individual abundance measurements from the five solar twins, while the colored symbols display the averages of these abundances calculated in linear space. The colored error bars are derived from error propagation based on the individual abundance errors. Fe is not included in these figures as it serves as a reference element, and K is absent from the GCE-corrected figure due to the lack of GCE correction for K. We perform a similar linear regression fitting including error bars using emcee, using the mean differential abundance from the five solar twins (colored symbols in Figure \ref{fig:differential}). The expression for the linear regression is -6.726e-05$_{-2.342e-05}^{+2.320e-05}$ $\times$ $T_c$ (K) + 0.086$_{-0.030}^{+0.030}$. The negative slope is significant at the 2$\sigma$ level, suggesting that the Sun is depleted in refractory elements relative to volatile elements when compared to the average abundance of the five solar twins.
	
	For the GCE-corrected abundances of our solar twins, the average [X/Fe]$_{\rm solar}$ - [X/Fe]$_{\rm star}$ for volatiles is 0.023 $\pm$ 0.003 dex, and for refractories, it is -0.005 $\pm$ 0.003 dex. This suggests an over-abundance ratio of volatiles to refractories in the Sun. We also notice that the Sun's overabundance of volatiles is primarily due to carbon and oxygen. After excluding these two elements, the average [X/Fe]$_{\rm solar}$ - [X/Fe]$_{\rm star}$ for the remaining volatile elements is -0.018 $\pm$ 0.003 dex. Based on our current analysis of the five solar twins, the Sun appears to be especially more enriched in C and O than these twins. When excluding C and O, the Sun shows a slight deficiency in both volatile and refractory elements compared to the five solar twins. When excluding C and O, the $T_c$ trend slope is 3.052e-05$_{-0.0001}^{+0.0001}$.
	
	\section{Discussions}  \label{sec:discussion}
	
	Our findings agree with previous studies that the Sun is relatively depleted in refractory elements compared to nearby field solar twins (e.g., \citealt{2024ApJ...965..176R}). However, the underlying cause of this deficiency remains under debate. We discuss hereafter the proposed causes of this trend, i.e., galactic chemical evolution and planet formation, respectively. In our discussion, we classify all planets with R $<$ 7 R$\bigoplus$ as terrestrial planets and those with R $\geq$ 7 R$\bigoplus$ as giant planets.
	
	The formation of terrestrial planets has been suggested as a possible mechanism for this depletion. Refractory materials in the Solar protoplanetary disk may have been locked into planetesimals before the Sun’s accretion phase (\citealt{2009ApJ...704L..66M}), leading to the depletion of refractory elements. \citet{2010ApJ...724...92C} supported this idea by showing that adding 4 Earth masses of terrestrial material to the Sun’s convection zone would eliminate the observed abundance differences. However, this explanation remains debated.
	
	Giant planet formation has also been proposed as a possible explanation. A forming giant planet can open a gap in the gas disk, creating a pressure trap. If the planet forms very early, while the disk is still dense, it could trap over 100 Earth masses of dust outside its orbit, preventing this material from accreting onto the star. This process could result in refractory element depletions of $\sim$5–15\% in the host star (\citealt{2020MNRAS.493.5079B}). However, no observational evidence has yet supported this mechanism, partly due to the limited availability of high-resolution and S/N spectra for solar twins hosting giant planets.
	
	\subsection{The stellar evolutionary effect on $T_c$ trend slopes}
	
\begin{figure*}[!ht]
		\centering			
		\includegraphics[width=0.6\textwidth, trim={5cm 0 5cm 0}]{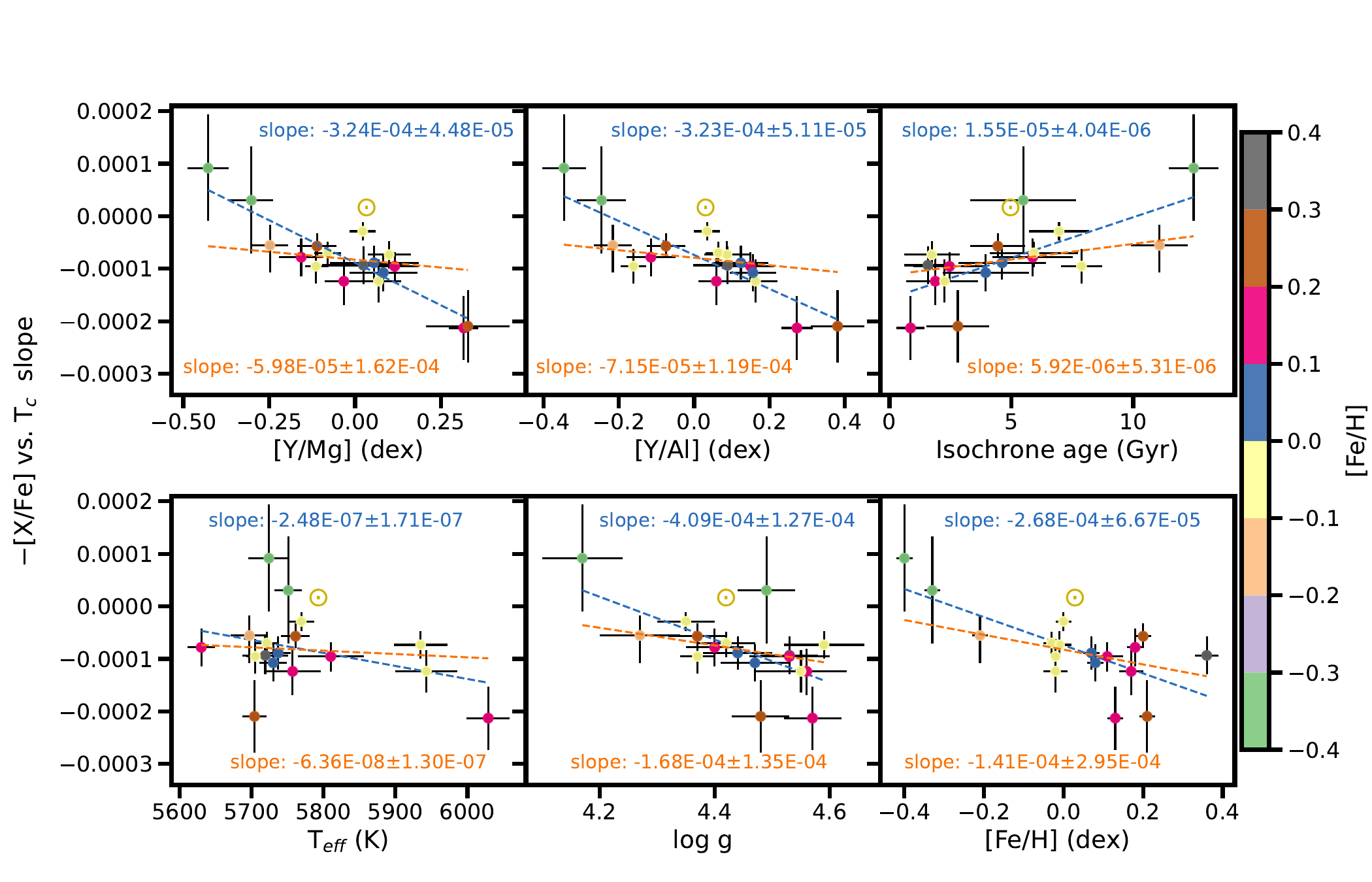}
		\caption{The $T_c$ trend slopes versus multiple stellar parameters, color-scaled by [Fe/H]. The parameters include [Y/Mg], [Al/Mg], isochrone age, $T_{\rm eff}$, log g, and [Fe/H]. The blue dashed line represents the best linear fit for all stars, while the orange dashed line indicates the best-fit linear regression for stars with $\Delta [Fe/H] < 0.1$ dex compared to the Sun. The slopes and associated errors are displayed in the plots: blue text denotes the slope and errors for all stars, and orange text indicates the slope and errors for the restricted sample. Although a significant linear trend is observed across all stars, it becomes insignificant at the 2-$\sigma$ level when focusing on the more uniform sample with $\Delta [Fe/H] < 0.1$ dex. The position of the Sun is marked with a yellow $\odot$ symbol.}
		\label{fig:Tc_param}
	\end{figure*}

	We first examine whether the $T_c$ trend slope results from stellar evolutionary effects. Only after ruling out stellar effects can we consider the impact of planet formation. Figure \ref{fig:Tc_param} shows the $T_c$ trend slope against various stellar parameters, including [Y/Mg], [Al/Mg], isochrone age, $T_{\rm eff}$, log g, and [Fe/H]. We aim to investigate whether the $T_c$ trend is potentially driven by stellar properties. We perform linear regression fits for each $T_c$ versus stellar parameter subplot, shown by the blue dashed lines. After restricting to stars with a difference in [Fe/H] of less than 0.1 dex from the Sun, we apply similar linear fittings, shown as orange lines. The slopes and 1-$\sigma$ errors are indicated in blue and orange text, respectively, on the subplots. The $T_c$ trends are significant across the range -0.4 $<$ [Fe/H] $<$ 0.4 dex for all 17 stars; however, when restricting $|$[Fe/H]$|$ to 0.1 dex, all trends become insignificant at the 2-$\sigma$ level. This is mainly due to galactic chemical evolution, which is significant for stars with varying metallicity. When restricting $|$[Fe/H]$|$ to 0.1 dex, the GCE effect remains but is less significant.
	
\begin{figure*}[!ht]
		\centering			
		\includegraphics[width=0.5\textwidth, trim={5cm 0 5cm 0}]{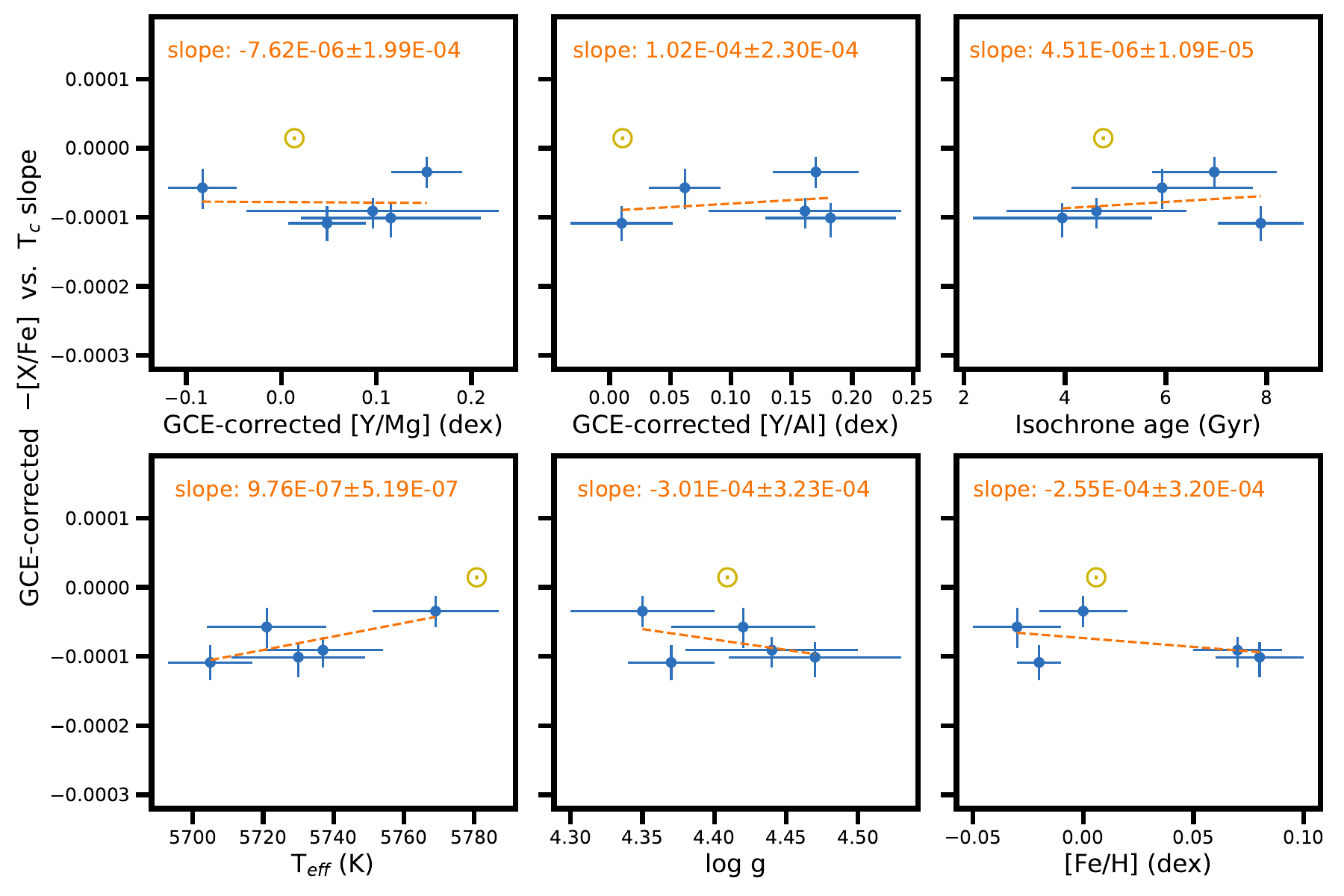}
		\caption{The [X/Fe]$_{\rm solar}$ - [X/Fe]$_{\rm star}$ (equivalent to $-$[X/Fe]) vs. $T_c$ trend slopes for GCE-corrected abundances of five solar twins (TOI-1055, TOI-2426, TOI-744, TOI-755, TOI-818) are plotted against various stellar parameters, color-scaled by [Fe/H]. The parameters include [Y/Mg], [Al/Mg], isochrone age, $T_{\rm eff}$, log g, and [Fe/H], consistent with those in Figure \ref{fig:Tc_param}. We fit a linear regression to each of the subplot, shown as an orange dashed line, with the slope and errors shown in orange text. In all cases, the slope is not significant at the 2-$\sigma$ level. The position of the Sun is marked with a yellow $\odot$ symbol.}
		\label{fig:Tc_GCE_param}
	\end{figure*}

	In Figure \ref{fig:Tc_GCE_param}, we present the $T_c$ trend slopes for the five solar twins with GCE-corrected abundances, using the same stellar parameters on the x-axis as those in Figure \ref{fig:Tc_GCE_param}. Linear regression fittings are performed for each subplot, with slopes and errors displayed accordingly. We find that the $T_c$ trend slopes show no dependence on the examined stellar parameters, including [Y/Mg], [Al/Mg], isochrone age, $T_{\rm eff}$, log g, and [Fe/H]. [Y/Mg] and [Al/Mg] serve as proxies for age (e.g., \citealt{2017A&A...604L...8S}). We plan to derive our own [Y/Mg] – age relation across all metallicities at the conclusion of this survey. This examination allows us to rule out GCE as the main cause of the Sun’s Tc trend.
	
	\subsection{Impact of Terrestrial and Giant Planet Formation on $T_c$ Trend Slopes}
	
	We now explore whether planet formation is related to the $T_c$ trend slopes and the depletion of refractory elements in the Sun.
	
	We first examine planetary parameters, including whether the stars host terrestrial and/or giant planets, the number of planets they are known to host, and the total rocky planetary mass. The formation of terrestrial and/or giant planets potentially leaves imprints on the host stars' compositions of volatile and refractory elements.
	
\begin{figure*}[!ht]
		\centering
		\includegraphics[width=0.5\textwidth, trim={5cm 0 5cm 0}]{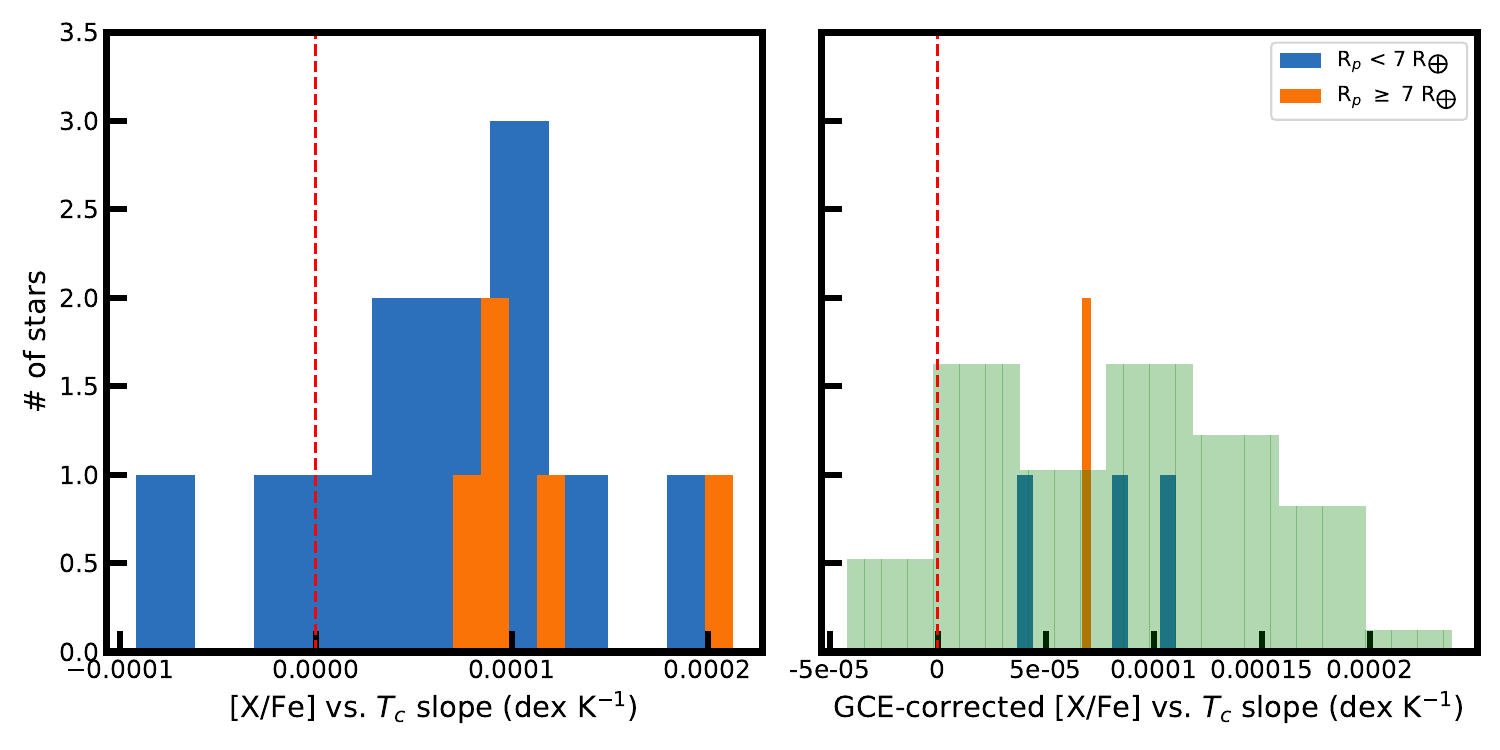}
		\caption{The [X/Fe]$_{\rm star}$ - [X/Fe]$_{\rm solar}$ (equivalent to [X/Fe]) vs. $T_c$ trend slope distribution for stars with and without giant planets. The left panel displays the [X/Fe] vs. $T_c$ trend slope distribution for all 17 Sun-like stars, while the right panel shows the GCE-corrected [X/Fe] vs. $T_c$ trend slope distributions for five solar twins. The stars hosting giant planets ($R_p$ $\geq$ 7 R$_\bigoplus$) are shown in orange, while those only hosting small planets ($R_p$ $<$ 7 R$_\bigoplus$) are shown in blue. The Sun, with a $T_c$ trend slope of 0, is plotted as a reference. The solar twin distributions from \citet{Bedell2018} are shown by the green histogram in the background.}
		\label{fig:Tc_dist}
	\end{figure*}

	We compare the $T_c$ slopes of systems with small and large planets to investigate whether giant planet formation is the dominant cause of the Sun's refractory element depletion. In Figure \ref{fig:Tc_dist}, the left panel compares the [X/Fe] vs. $T_c$ trend slope distribution for 17 solar-like stars, distinguishing those with and without giant planets (R$p$ $\geq$ 7 R${\bigoplus}$). The right panel focuses on the [X/Fe] vs. $T_c$ trend slopes of the five solar twins with GCE-corrected abundances. In both panels, we have reversed the slope of $-$[X/Fe] vs. $T_c$ to [X/Fe] vs. $T_c$, allowing for a direct comparison with the [X/Fe] vs. $T_c$ trend distributions of 68 solar twins from \citet{Bedell2018} (top panel of their Figure 8). In the right panel, we overlay the GCE-corrected sample from \citet{Bedell2018} as a background distribution (green histogram). Since we only have 5 solar twins, we divide the distribution from \citet{Bedell2018} by 10 to facilitate a direct comparison. Our $T_c$ trend distribution of solar twins agree with theirs. In both panels, all stars hosting giant planets have [X/Fe] vs. $T_c$ trend slopes above zero, indicating that these stars are normal and undepleted, similar to field solar twins. The Sun, which is depleted in refractory elements, has a $T_c$ slope of zero.
	
\begin{figure*}[!ht]
		\centering
		\includegraphics[width=0.5\textwidth, trim={5cm 0 5cm 0}]{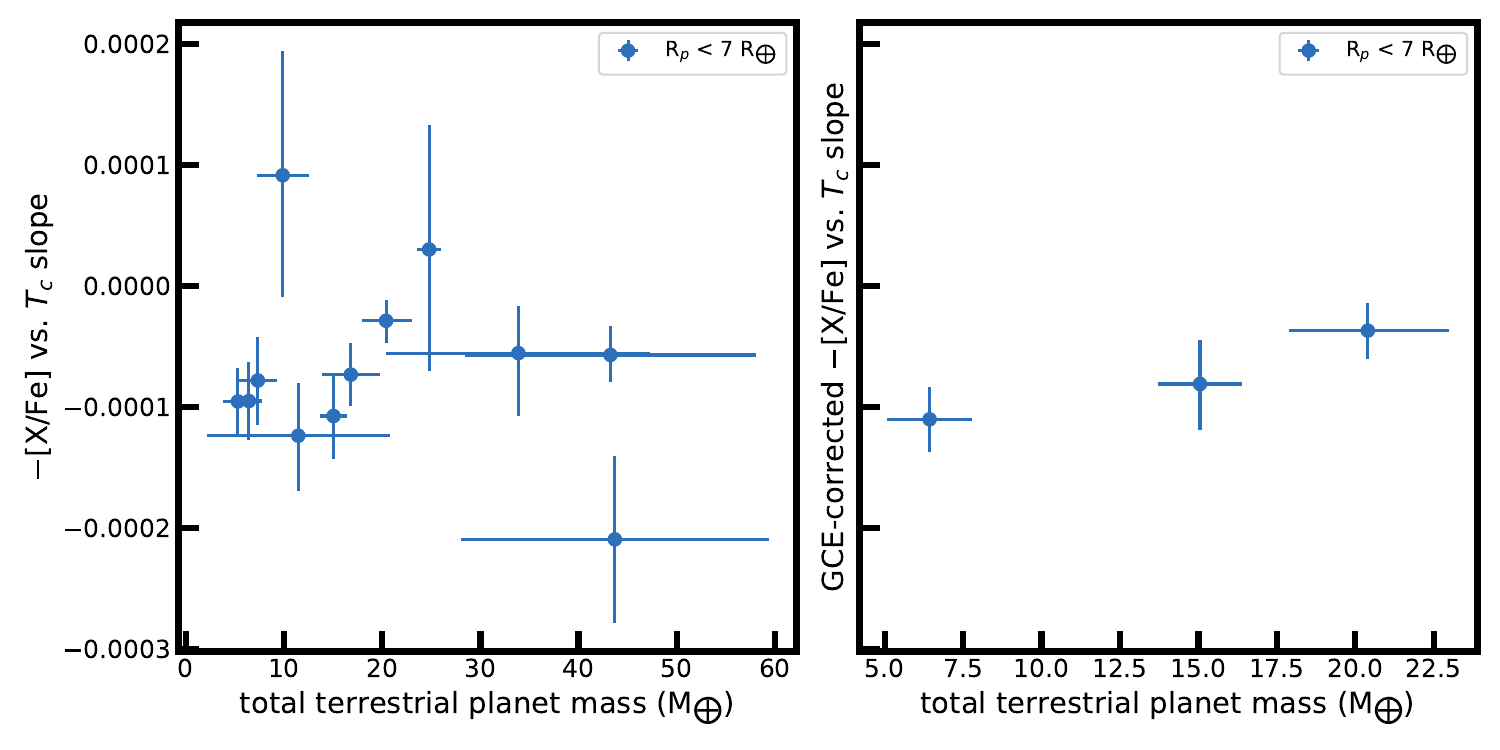}
		\caption{The $T_c$ trend slope versus total terrestrial planet mass. The left panel shows all solar-like stars hosting terrestrial planets, while the right panel focuses on the three solar twins hosting terrestrial planets. We compute planet mass by using the planet radius in Table \ref{tab:parameters}, and the mass-radius relation for small planets from the Spright Python package (\citealt{2024MNRAS.527.5693P}).}
		\label{fig:Tc_totmass}
	\end{figure*}

	We then examine whether terrestrial planet formation affects the $T_c$ trend slope distributions. In Figure  \ref{fig:Tc_totmass}, we show the $T_c$ trend slope distribution as a function of the total terrestrial planet mass. We calculate planet mass using planet radii from Table \ref{tab:parameters} and the mass-radius relation for small planets provided by the Spright Python package (\citealt{2024MNRAS.527.5693P}). If terrestrial planet formation indeed leads to the refractory element depletions in the Sun, we would expect to observe a more negative [X/Fe]$_{\rm solar}$ - [X/Fe]$_{\rm star}$ vs. $T_c$ trend for stars hosting more rocky planets. However, we observe no correlation between $T_c$ trend slopes and the total terrestrial planet mass. This is consistent with \cite{2021MNRAS.507.2220G}, who found that solar analogs with rocky planets do not show refractory element depletion like the Sun. Our analysis suggests that small planets may not be responsible for the Sun's refractory element depletion. Moreover, the 17 solar-like stars with small planets mostly have negative $T_c$ trends, further suggesting that rocky planet formation is unlikely the cause.
	
	We remind readers that only transiting planets are detected, which may lead to some planets being omitted. Additionally, stars with inner small planets may host undetected outer cold Jupiters missed by TESS (\citealt{2018ApJ...860..101Z}). If a star hosts undetected planets or outer cold Jupiters missed by TESS, we may be underestimating the total terrestrial planet mass. The cores of giant planets could provide an additional source of rocky material beyond the terrestrial planet mass. This would increase the actual total rocky mass and potentially create a correlation between the $T_c$ trend slopes and the total planet mass in Figure \ref{fig:Tc_totmass}. However, we are currently unable to resolve this through observations.
	
	Even in the Solar System, where the number and mass of terrestrial and giant planets are well known, the ice-to-rock composition of the giant planets remains uncertain. As \citet{2018A&A...618A.132K} suggested, the total planetary mass in the Solar System relative to the total mass not in hydrogen and helium is estimated to be between 52 and 109 $M_{\bigoplus}$ for the present-day mass, and between 97 and 168 $M_{\bigoplus}$ when accounting for ejected mass. However, the precise ice-to-rock ratio in giant planets remains unknown (\citealt{2023ASPC..534..947G}).
	
	A larger sample could help distinguish these populations. We acknowledge that our current analysis is limited by the relatively small sample size. Future observing runs of the PASTA survey will focus on expanding the sample to address these questions.
	
	\section{Summary}   \label{sec:summary}
	
	This is Paper I of the PASTA: Planets Around Solar Twins/Analogs survey. In this first paper, we obtain high-resolution MIKE spectra for 17 solar-like stars that host or likely host planets. We report high-precision stellar abundances for 22 elements (from C to Eu) for the 17 planet-hosting stars, and achieve a 1-$\sigma$ error to 0.01 for elements such as Fe and Si, to 0.08 for elements like Sr, Y, and Eu. The 17 stars span −0.4 $<$ [Fe/H] $<$ 0.4 dex, 5630 $<$ $T_{\rm eff}$ $<$ 6029 K, and 4.17 $<$ log {\it g} $<$ 4.59. The elemental abundances in our study are in agreement with previous results from large surveys, such as GALAH. The variations in elemental abundance due to Galactic chemical evolution (GCE) effects are similar to those found in previous studies.
	
	For all the 17 solar-like stars, we compare their differential abundance to the Sun ([X/Fe]$_{\rm solar}$ - [X/Fe]$_{\rm star}$) and derive the differential abundance versus condensation temperature ($T_c$) trend. Our current sample of 17 stars, covering a range of metallicities around solar metallicity, includes five solar twins ($\Delta$[Fe/H] $<$ 0.1 dex, $\Delta\ T_{\rm eff}$ $<$ 100 K, $\Delta$ log g $<$ 0.1) as defined by \cite{Bedell2018}. After applying GCE corrections using relation from \cite{Bedell2018}, we perform similar $T_c$ trend analysis for the five solar twins.
	
	By restricting to the five solar twins only, we compute their average [X/Fe] abundance and compare to that of the Sun. We find a negative slope in the differential abundance ([X/Fe]$_{\rm solar}$ - [X/Fe]$_{\rm star}$) versus $T_c$ plot, suggesting that the Sun is relatively depleted in refractory elements, though the trend is primarily driven by carbon and oxygen. Our finding aligns with some previous studies suggesting that the Sun is depleted in refractory elements (e.g., \citealt{2024arXiv240216954R, 2021ApJ...907..116N}). At this stage, we acknowledge that the conclusion may be limited by small number statistics.
	
	We find that the $T_c$ trend slopes for the solar metallicity stars (-0.1 $<$ [Fe/H] $<$ 0.1 dex) with intermediate ages do not depend on the multiple stellar parameters, including [Y/Mg], stellar age, $T_{\rm eff}$, log g, [Fe/H], and microturbulence. The stars hosting giant planets all have negative [X/Fe]$_{\rm solar}$ - [X/Fe]$_{\rm star}$ vs. $T_c$ trend slopes, derived from both the original and GCE-corrected abundances. This suggests that the Sun, unlike other giant-planet host solar twins, is relatively depleted in refractory elements compared to volatile elements. Our results remain robust regardless of whether the GCE correction is applied. Furthermore, we observe no correlation between $T_c$ trend slopes and total terrestrial planet mass, indicating that terrestrial planet formation may not be responsible for the depletion of refractory elements in these stars. However, this analysis is based on a small sample size of 17 planet-hosting stars, so we acknowledge that our findings may be influenced by small number statistics. In addition, we note that our trend in Figure \ref{fig:Tc_trend}, \ref{fig:Tc_GCE_trend}, and \ref{fig:differential} are primarily driven by the abundances of carbon (C) and oxygen (O). We plan to revisit these issues as we observe more solar-like stars in future runs of the PASTA survey.
	
	In future runs of this survey, we will continue to observe solar-like stars using high-precision spectroscopy, with a primary focus on likely solar twins. Our PASTA target list will not only contain planet hosts discovered by TESS, but also Kepler or RV-discovered ones. This expanded dataset will be valuable in exploring whether the chemical composition of the Sun is influenced by the diverse stellar and planetary factors, such as the formation of giant and terrestrial planets. For example, comparing the $T_c$ trend slope distribution of solar twins with and without giant planets, and the $T_c$ trend slope versus terrestrial planet mass, will help us understand if planet formation affects the volatile and refractory elemental abundance of the host star.
	
\section*{acknowledgements}

Q.S. is supported by the National Key R\&D Program of China, No. 2024YFA1611800. Y.S.T. is supported by the National Science Foundation under Grant No. AST-240672. The postdoctoral fellowship of K.B. is funded by F.R.S.-FNRS grant T.0109.20 and by the Francqui Foundation.
	
This paper includes data gathered with the 6.5m Magellan Telescopes located at Las Campanas Observatory, Chile, kindly supported by Carnegie Observatories. Some of the targets in this work were observed with the ASTEP telescope. ASTEP benefited from the support of the French and Italian polar agencies IPEV and PNRA in the framework of the Concordia station program and from OCA, INSU, ESA, ERC (grant agreement No. 803193/BEBOP) and STFC; (grant No. ST/S00193X/1).
	
This paper made use of data collected by the TESS mission and are publicly available from the Mikulski Archive for Space Telescopes (MAST) operated by the Space Telescope Science Institute (STScI). Funding for the TESS mission is provided by NASA’s Science Mission Directorate. We acknowledge the use of public TESS data from pipelines at the TESS Science Office and at the TESS Science Processing Operations Center. Resources supporting this work were provided by the NASA High-End Computing (HEC) Program through the NASA Advanced Supercomputing (NAS) Division at Ames Research Center for the production of the SPOC data products.

	\appendix
	
	\section{The full sample list for the PASTA survey}
	
	The full sample for the PASTA survey is shown in Appendix Table A. It includes 188 stars, selected from the TESS Objects of Interest Catalog, including solar-like stars that host known or confirmed planets or planet candidates. 
	
	\begin{deluxetable*}{ccccccccccccccc}[!bp]
		\renewcommand\thetable{A}
		\tablecaption{The full sample list for the PASTA survey}
		\tabletypesize{\scriptsize}
		\tablehead{
			TIC ID & TOI ID & RA & Dec & Disposition	& $V$	& $R_p$	& $M_p$	& Period & Rs	& Ms	& $T_{\rm eff}$	& log g & facilities & status \\
			&  & $^{\circ}$ & $^{\circ}$  &	& mag	& $R_{\bigoplus}$	& $M_{\bigoplus}$& day	& $R_{\odot}$	& $M_{\bigoplus}$	& K	&  &  & 
		} 
		\startdata
		\hline
		4598935 &	12.35	&   141.220019	&  -39.030905	& PC	& 12.02	& 6.9	& --	    & 24.7	& 1.18	& 1.06	& 5876	& 4.32	& --	& --                  \\
		183301671 &	25.11	&   17.995374	&  -77.738828	& PC	& 12.65	& 3.3	& --	    & 0.6	& 1.10	& 1.06	& 5863	& 4.38	& --	& --                  \\
		... &	...	&   ...	&  ...	& ...	& ...	& ...	& ...	    & ...	& ...	& ...	& ...	& ... & ...	& ... \\
		\hline
		\enddata
		\tablecomments{1. Stellar and planetary parameters, including TESS Input Catalog (TIC) ID, TESS Object of Interest (TOI) Catalog ID, right ascension (RA), declination (DEC), TOI disposition, $V$ magnitude, planet radius (in Earth radii), orbital period, host star radius, stellar mass, effective temperature ($T_{\rm eff}$), surface gravity (log {\it g}), observing facilities (if any), and current TESS status. \\
			(This table is available in its entirety in machine-readable form online.)}
		\vspace{-10mm}
	\end{deluxetable*}
	
	\bibliography{sun24_solar}{}
	
	\bibliographystyle{aasjournal}
	
\end{document}